\documentstyle[epsf]{mn}
\newcommand{\lsim}{\raisebox{-0.6ex}{$\stackrel{{\displaystyle<}}{\sim}$}}
\newcommand{\gsim}{\raisebox{-0.6ex}{$\stackrel{{\displaystyle>}}{\sim}$}}
\newcommand{\doverd}[2]{\frac{\partial #1}{\partial #2}}
\newcommand{\doverdt}[1]{\frac{\partial #1}{\partial t}}

\newcommand{\subscr}[1]{_{\rm #1}}

\newcommand{\Mdot}{\dot{M}}
\newcommand{\Msol}{\rm M_{\odot}}

\newcommand{\bequ}{\begin{equation}}
\newcommand{\eequ}{\end{equation}}

\title{Mass Flow and Accretion through gaps in Accretion Discs}
\author[W. Kley]
       {W. Kley\\
       Theoretisch-Physikalisches Institut,
       Universit\"at Jena, Max-Wien-Platz~1, D-07743 Jena, Germany\\
       }
\begin{document}
\maketitle
\epsfverbosetrue
\begin{abstract}
We study the structure and dynamics of the gap created by a protoplanet in
an accretion disc. The hydrodynamic equations for a
flat, two-dimensional, non-selfgravitating protostellar accretion disc
with an embedded, Jupiter sized protoplanet on a circular orbit are solved.
To simulate possible accretion of mass onto the protoplanet we continually
remove mass from the interior of the planet's Roche lobe which is monitored.
Firstly, it is shown that consistent results independent on numerical issues
(such as boundary or initial conditions, artificial viscosity or resolution)
can be obtained. Then, a detailed parameter study delineates the influence
of the disc viscosity and pressure on the magnitude of the accretion rate.

We find that, even after the formation of a gap
in the disc, the planet is still able to accrete more mass from the disc.
This accretion occurs from regions of the disc which are radially
exterior and interior to the planet's orbital radius.
The rate depends on the magnitude of the viscosity and
vertical thickness of the disc.
For a disc viscosity $\alpha=10^{-3}$ and vertical thickness
$H/r=0.05$ we estimate the time scale for the accumulation
of one Jupiter mass to be of order hundred thousand years.
For a larger(smaller) viscosity and disc thickness this accretion
rate is increasing(decreasing).

For a very small viscosity $\alpha \lsim 5 \, 10^{-4}$ the mass accretion rate
through the gap onto the planet is markedly reduced, and the corresponding
accretion time scale becomes larger than the viscous evolution time of
the disc.
\end{abstract}
\section{Introduction}
In standard formation theories of giant (Jupiter sized) planets it is
assumed that after the initial buildup of a central rocky core
of a few earth masses
further mass growth proceeds through accretion of gas from the surrounding
accretion disc (Boss 1996).
When sufficiently grown, the protoplanet exerts tidal torques on the
disc and induces trailing spiral shocks (Lin \& Papaloizou 1980,
Goldreich \& Tremaine 1980, and Papaloizou \& Lin 1984). These
transport angular momentum and material is pushed away from the protoplanet,
a process which leads eventually to the opening of a gap in the disc.
The detailed criteria of gap opening are given
by Lin \& Papaloizou (1986, 1993). Viscous torques in the disc
counteract the gravitational torques generated by the protoplanet,
leading to an equilibrium configuration. The radial extent of this gap
depends on the mass of the protoplanet, the magnitude of the viscosity  
and the pressure in the disc
(Lin \& Papaloizou 1993; and references therein).
For solar system parameter of the protostellar
disc one finds that the critical mass for gap opening to occur
is of the order of about one Jupiter mass $M_J$ (Lin \& Papaloizou 1993).
 
Usually, it has been assumed that after the gap has been formed further mass
growth of the planet is inhibited, which then yields a natural limit for
the final mass of the planet. However, numerical calculations of discs
around a binary star by Artymowicz \& Lubow (1996) have
shown that, even after a gap has formed, further accretion through the
gap onto the two stars may nevertheless take place.
They argue that a similar (gap accretion) process may also occur
in the case of a protoplanet embedded in a disc.

This process of the interaction of the protoplanet with its
surrounding disc has recently attracted much attention (eg. Glanz 1997)
after the discovery of extrasolar planets around solar type stars
(Mayor \& Queloz 1995; Butler \& Marcy 1996; Marcy \& Butler 1996).
Their properties are summarized by Boss (1996) and Cochran (1997).
They fall basically into two groups: One with very low ($e \lsim 0.1$)
eccentricities and minimum masses in the range 0.47 to 3.66 $M_J$,
and the second with three planets having masses between
1.67 and 10 $M_J$ and eccentricities of $e \gsim 0.4$
(Mazeh, Mayor \& Latham 1997;
Marcy \& Butler 1998). The question arises whether it is possible to
form planets with masses of several $M_J$ by continued accretion
through the gap. 

The stationary problem of a planet in a disc has been analyzed by
Miki (1982) and more recently by Korycansky \& Papaloizou (1996). They
find that indeed the planet's
potential creates trailing spiral waves in the disc leading all the way down
to the planet. However due to the local shearing sheet and stationarity
assumption global properties of the spiral arm and possible mass accretion
onto the planet could not be analysed. These calculations nevertheless
indicated that hydrodynamic accretion onto a protoplanet from a Keplerian
disc results in a strong prograde rotation for the planet, resolving an
old problem in planet formation theory quite naturally (Hoyle 1946).
Time dependent calculations by Sekiya, Miyama \& Hayashi (1988)
showed that the trailing arms are a global phenomenon. Additionally
they found the existence of multiple arms inside the planet's orbit.
They showed that for a Jupiter sized planet a gap is opened in the disc
but did not see any indication of mass accretion through this gap.

Here we present a more elaborate study of the feasibility of this process by
performing numerical calculations of a thin, non-selfgravitating,
viscous disc with an embedded protoplanet.
The planet is assumed to be on a circular orbit with vanishing eccentricity.
We run the models for 400 to 1000 orbital periods of the planet,
and investigate in detail the effects of viscosity and pressure in the disc.
A similar investigation has been carried out by Bryden et al. (1998).

In the next section we present the physical
model, describe briefly the numerical method applied, and present a simple
test problem for the viscosity.
In section 3, we first
carefully analyse possible numerical effects, and
come to the conclusion that the inferred mass transfer rate is
independent of numerical issues.
We then proceed to study the influence of a variation of viscosity
and temperature in the disc on the mass accretion rate.
Our conclusions are given in Section 4. 

%%-- Sect.2
\section{Physical Model}
\noindent 
\subsection{Basic Equations}
In an accretion disc the vertical
thickness $H$ is usually assumed to be small in comparison to the distance $r$
from the centre, i.e. $H/r << 1$. This is naturally expected when the
material is in a state of near Keplerian rotation. Then one can
vertically integrate the hydrodynamical equations and  work only with
vertically averaged state variables. 

To achieve an increased accuracy of the flow in the vicinity
of the protoplanet, it will be convenient during the computations
to have the protoplanet
at a fixed location in the grid. Thus, we will work in a coordinate system
which corotates with the orbital angular velocity of the planet.
In a reference frame rotating with any (constant)
angular velocity vector ${\bf \Omega}$,
the total time derivative of the velocity ${\bf u}$ is given by
\begin{equation}
  \frac{d {\bf u}}{dt}  \left( =
  \doverd{{\bf u}}{t} + {\bf u} \nabla {\bf u} \right) =
     - 2 \, {\bf \Omega} \times {\bf u}  
     - {\bf \Omega} \times ( {\bf \Omega} \times {\bf r} ),
\end{equation}
where the first term on the right hand side describes Coriolis and
the second the centrifugal accelerations created by the rotating reference
frame.

We shall work in cylindrical coordinates ($r, \varphi, z$),
where $r$ is the radial coordinate, $\varphi$ is the azimuthal angle,
and $z$ is the vertical axis, and the rotation is around the $z$-axis, i.e
${\bf \Omega} = (0,0,\Omega)$. The origin of the
coordinate system is located at the centre of mass of the
system. 
In a flat disc located
in the $z=0$ plane, the
velocity components are ${\bf u} = (u_r, u_{\varphi}, 0)$.
In the following we will use the symbol $v=u_r$ for the radial velocity
and $\omega = u_{\varphi}/r$ for the angular velocity of the flow,
which are measured in the corotating frame.
Then the vertically integrated equations of motion are
%%  Continuity
\begin {equation}
 \doverdt{\Sigma} + \nabla (\Sigma {\bf u} )
                =  0,  \label{Sigma}
\end{equation}
%%  Radial Momentum
\begin {equation}
 \doverdt{(\Sigma v)} + \nabla (\Sigma v {\bf u} )
  = \Sigma \, r ( \omega + \Omega)^2
        - \doverd{p}{r} - \Sigma \doverd{\Phi}{r} + f\subscr{r}
      \label{u_r}
\end{equation}
%%  Azimuthal (Angular) Momentum
\begin {equation}
 \doverd{[\Sigma r^2 (\omega + \Omega)]}{t} 
   + \nabla [\Sigma r^2 (\omega + \Omega) {\bf u} ]
         =
        - \doverd{p}{\varphi} - \Sigma \doverd{\Phi}{\varphi}
   + f\subscr{\varphi}
      \label{u_phi}
\end{equation}
Here $\Sigma$ denotes the surface density
\[  \Sigma = \int^\infty_{-\infty} \rho dz,\]
where $\rho$ is the density, $p$ the
vertically integrated (two-dimensional) pressure. The
gravitational potential $\Phi$ created by the protostar
with mass $M_s$ and the planet having mass $m_p$ is given by
\[
    \Phi = - \frac{G M_s}{| {\bf r} - {\bf r}_s |}
       -  \frac{G m_p}{| {\bf r} - {\bf r}_p |},
\]
where $G$ is the gravitational constant and ${\bf r}_s,
{\bf r}_p$ are the radius vectors to the star and planet, respectively.
The effects of viscosity are contained in the terms $f_r$, and $f_{\varphi}$
which give the viscous force per unit area acting
in the radial and azimuthal direction. The explicit form of these terms is
described below.

Since the mass of the planet is very small in comparison to the mass
of the star, we use here a ratio $q = m_p/M_s = 10^{-3}$,
the centre of mass is located
very closely to the position of the star. In the following we will
frequently refer to parameters (such as temperature) of the
unperturbed disc (i.e. no planet) as a function of radius.
For simplicity we will identify in those cases the radial distance
from the origin with the distance from the central star.

\subsection{Equation of state}
In the set of equations above we have omitted the energy equation
because in this study we will be concerned only with relatively
simple equations of state which do not require the solution of
an energy equation.
Primarily, we will use an isothermal equation of state
where the surface pressure $p$ is related to the density $\Sigma$ and  
temperature $T$ through
\begin{equation}
        p = \Sigma \, c\subscr{s}^2.
\end{equation}
The local isothermal sound speed $c\subscr{s}$ is given here
by
\bequ     \label{cs}
        c\subscr{s} = \frac{H}{r} \, v\subscr{Kep},
\eequ
where $v\subscr{Kep} = \sqrt{G M_s/r}$ denotes the Keplerian orbital velocity 
of the unperturbed disc.
Equation (\ref{cs}) follows from vertical hydrostatic equilibrium.
The ratio $h\subscr{r}$ of the vertical height $H$ to the
radial distance $r$ is taken as a fixed input parameter. Here we use a
standard value of
\[      h\subscr{r} =  H/r =  0.05  \label{hr}, \]
which is typical for protostellar accretion discs having a mass
inflow rate of $\Mdot \approx 10^{-7} \Msol/yr$.

Alternatively, to explore further parameter space,
we will also use a polytropic equation of state
\bequ  \label{poly}
       p = K \, \Sigma^\gamma
\eequ
with the constant $K$ and
the adiabatic exponent $\gamma = 2$.
In the polytropic equation of state (\ref{poly}) the sound speed
is given by $c_s^2 = \gamma p/\Sigma$ and then, 
according to (\ref{cs}), the ratio $H/r$ is radius dependent.
The constant $K$ is adjusted in such a way to make
the ratio $h_r = 0.05$ at the planetary radius $r_p$ in an unperturbed
disc.

The prime difference between an isothermal and polytropic disc is
the behaviour in the gap region in the vicinity of the planet. In the
first, isothermal case a fixed value of $h_r$ is specified, hence
the temperature in the disc has the radial profile
$T(r) \propto r^{-1}$, not influenced by the density, while
in the second polytropic case the temperature is reduced in
the gap region. Thus, the polytropic equation could mimic some
cooling processes in optically thin regions.
This may possibly influence the magnitude of the viscosity
within the gap region.

\subsection{Viscosity}
Viscous processes are of central importance in accretion discs in that they
are responsible for the  angular momentum transport that allows for
radial inflow and accretion to occur. It is believed that processes such as MHD
turbulence are likely to be responsible for the existence of the
large viscosities required to account for observed evolutionary timescales
associated with protostellar discs
(see Papaloizou and Lin 1993, and references therein).
We assume that the global effect of the turbulence, whatever origin,
can be modeled by
Reynolds stresses, which can be cast into a form, mathematically
identical to the standard viscous stress tensor only with the
molecular viscosity replaced by a turbulent
(eddy) viscosity coefficient $\nu_t$.

Using this ansatz, the viscous terms in the equations of motion (\ref{u_r},
\ref{u_phi}) are given by
\bequ
      f\subscr{r} = \nabla \cdot {\bf S}\subscr{r} 
         - \frac{S\subscr{\varphi\varphi}}{r}    \label{f_r}
\eequ
\bequ
      f\subscr{\varphi} = \nabla \cdot 
      \left( r \, {\bf S}\subscr{\varphi} \right) \label{f_phi} 
\eequ
with the vectors ${\bf S}\subscr{i} 
= (S\subscr{ir},S\subscr{i\varphi},S\subscr{iz})$ and $(i = r, \varphi, z)$.
In case of a motion confined to the equatorial plane ($z=0, u\subscr{z}=0$)
the only relevant components of
the viscous stress tensor $S\subscr{ij}$ are
\begin{eqnarray}
   S\subscr{rr}  & = & 2 \eta \left( \frac{\partial v}{\partial r} 
          \right) 
     + \, \left( \zeta - \frac{2}{3}\eta \right) \nabla \cdot {\bf u}\\
   S\subscr{\varphi\varphi}  & = &
     2 \eta \left( \frac{\partial \omega}{\partial \varphi} 
          + \frac{v}{r} \right) 
     + \, \left( \zeta - \frac{2}{3}\eta \right) \nabla \cdot {\bf u}\\
   S\subscr{zz}  & = &
          \left( \zeta - \frac{2}{3}\eta \right) \nabla \cdot {\bf u}\\
   S\subscr{r\varphi}  & = &
      \eta \left( \frac{1}{r} \, \frac{\partial v}{\partial \varphi} 
          + r \frac{\partial \omega}{\partial r} \right)
\end{eqnarray}
where $\nabla \cdot {\bf u}$ denotes the divergence of the velocity
\[
      \nabla \cdot {\bf u} =  \frac{\partial v}{\partial r} + \frac{v}{r}
           + \frac{\partial \omega}{\partial \varphi},
\]
and $\eta$ and $\zeta$ are the shear and bulk viscosity, respectively.
As in standard accretion disc theory the physical
bulk viscosity $\zeta$ is assumed to be zero.
However, as described below, we will use a non-zero $\zeta$ for an
additional artificial viscosity.

For the shear viscosity coefficient we write $\eta = \rho \nu\subscr{t}$,
where $\nu\subscr{t}$ denotes the effective turbulent kinematic
viscosity.
For simplicity we use in our standard model a constant viscosity
\bequ    \label{nu_const}
        \nu\subscr{t} = const.
\eequ

In accretion disc theory the $\alpha$ prescription of Shakura and
Sunyaev (1973) is often used such that 
\begin{equation}
   \nu\subscr{t} = \alpha c\subscr{s} H.  \label{nu_alp}
\end{equation}
Here $\alpha$ is a (usually constant) coefficient of proportionality
describing the efficiency of the turbulent transport.
In writing and (\ref{nu_alp}) it is envisaged that the
turbulence behaves in such a way as to produce  a viscosity through the
action of eddies of typical size  $H$ and  turnover velocity
$\alpha c\subscr{s}$. This $\alpha$ prescription we will use
along with the constant viscosity for a parameter study.
One of the main difference of the two models is, again,
the behaviour in the gap. While the $\alpha$ viscosity can be very low in
the gap (for example in case of the polytropic equation of state),
in the first prescription (\ref{nu_const}), $\nu\subscr{t}$ is
of course unaffected.

\subsubsection{Artificial Viscosity}
Even though the problem has in general a rather large physical
(turbulent) viscosity, it may nevertheless be necessary to implement
an additional explicit artificial viscosity. This is particularly
required in regions with possibly lower mass density, as this may cause
unstable behaviour in the velocities.
The intrinsic numerical viscosity of the code in those cases may not be
sufficient to damp these.
Hence, an additional artificial viscosity is used during the
calculations, also to test its possible influence on the physics of
the problem. As will be described later in the results section, we found it
necessary to apply the artificial viscosity only to the bulk part
($\zeta$) of the viscosity.

To prevent any unphysical overshooting near steep gradients of the velocity
we use for the artificial kinematic viscosity coefficient
\bequ  \label{nu_art}
       \nu\subscr{a} = \left\{ \begin{array}{ll}
             - C\subscr{a} \delta^2 \, \nabla \cdot {\bf u} & 
                  \mbox{if $\nabla \cdot {\bf u} < 0$}  \\
                  0   &
                  \mbox{otherwise}
             \end{array}  \right.
\eequ
Here $\delta$ denotes the extension of the gridcell under consideration,
and the constant $C\subscr{a}$ is of order unity, which
is typically set to $0.5$ in the present work.
Then the bulk viscosity is given by $\zeta = \rho \nu\subscr{a}$.

\subsection{Numerical Considerations}

\subsubsection{Dimensionless Units}
For numerical convenience we introduce dimensionless units, in which the
distance $a$ of the planet to the star is taken as the unit of length,
$R_0=a$.
In physical units this could be for example the actual distance of
Jupiter from the sun $a\subscr{J}=5.2 AU$.
The unit of time is obtained from the orbital angular frequency 
$\Omega\subscr{p}$ of the planet
\bequ  \label{t_0} 
      t_0 = \Omega\subscr{p}^{-1} =  
         \left( \frac{a^3}{G (M\subscr{s} + m\subscr{p})} \right)^{1/2},
\eequ
i.e. the orbital period of the planet is given by 
\bequ
   P_p = 2 \, \pi t_0  \label{Pp}.
\eequ
The evolutionary time of the results of the calculations as given below
will usually be stated in units of $P_p$. The unit of velocity is then
given by $v_0 = R_0 / t_0$.
Equation (\ref{t_0}) could be also viewed as normalizing the unit of mass 
to the total mass of the system $M\subscr{t} = M_s + m_p$,
giving that the gravitational constant $G$ will be set to one, as is done
frequently in relativistic calculations. Since the unit of density
$\Sigma_0$ drops out of the equations of motion, we may
normalize it to any specified density. It will be useful to
have the total mass in the disc be adjusted to a given fraction 
(eg. $10^{-2}$) of the stellar mass.
Then the accreted mass onto the planet can be compared to the planet's
initial mass.
The unit of the kinematic viscosity coefficient is given by
$\nu_0 = R_0 v_0$, and the constant value of the viscosity is given in
units of $\nu_0$.

\subsubsection{The numerical method in brief}
The normalized equations of motion (\ref{Sigma} - \ref{u_phi}) are
solved using an Eulerian finite difference scheme,
where the computational domain $[r\subscr{min}, r\subscr{max}] \times
[\varphi\subscr{min}, \varphi\subscr{max}]$
is subdivided into $N_r \times N_\varphi$
grid cells. For the typical runs we use $N_r = 128, N_\varphi = 128$,
where the azimuthal spacing is equidistant, and the radial
points have a closer spacing near the inner radius.
In the azimuthal direction a whole ring is computed.
The numerical method is based on a spatially second order accurate
upwind scheme (monotonic transport), which uses a formally first
order time-stepping procedure. The basic features are described
in Kley (1989), and we will give only a very brief summary of the
changes and additions here. 

The force and advection terms are solved explicitly, and for stability
reasons the usual Courant-Friedrich-Levy time step criterion has to be
applied
\bequ
        \Delta t = f\subscr{C} \, 
           \min_{ij} \left( \frac{\Delta r_i}{|v| + c_s}, \,
           \frac{\, \Delta \varphi_j}{|\Omega + \omega| + c_s/r} 
          \right)
\eequ
where $\Delta r_i$ and $\Delta \varphi_j$ denote the sizes of the
$i^{th}$ radial and $j^{th}$ azimuthal grid cell.
The Courant number $f\subscr{C}$ is set to 1/2.
To prevent numerical problems near the planet where the gravitational
potential $\Phi_p$ of the planet diverges,
we introduce a softening of the potential
in the following form
\bequ
    \Phi_p = - \frac{G m_p}
      {\left[ ({\bf r} - {\bf r}_p)^2 + r_{sm}^2 \right]^{1/2}}
\eequ
where $r_{sm}$ is a smoothing length defined by
\[
     r_{sm} = \frac{1}{5} \left( \frac{\mu}{3} \right)^{1/3},
\]
i.e. 1/5 of the approximate size of the Roche lobe of the planet.
The variable $\mu$ denotes the reduced mass $\mu = q / (1+q)$ of the system.
This smoothening of the potential allows a motion of the planet through
the grid, for example in the case of non-vanishing eccentricity or an
inertial frame $\Omega=0$ calculation.
The exact value of $r_{sm}$ is not so important as long as it is
substantially smaller than the Roche radius of the planet. Tests with different
values gave identical results.

It should be noted that the angular equation of motion (\ref{u_phi})
has been put into an explicitly conservative form,
indicating the conservation of angular momentum. Numerical experiments
have shown that in case of a non-zero
speed of the rotating reference frame ($\Omega \neq 0$) this conservative
form seems to be necessary for numerical stability, and it guarantees identical
results of inertial and rotating frame calculations (Kley 1998).
 
The contributions of the viscous terms (\ref{f_r}, \ref{f_phi}) are
solved implicitly, to avoid any further time step limitations.
This also includes the contributions of the artificial viscosity.
The resulting linear system of equations is solved by the method of
Successive Overrelaxation (SOR).

\subsubsection{Boundary and initial conditions} \label{bounds}
To cover the range of influence of the planet on the disc
fully, we typically chose for the
boundaries (in dimensionless units, where the planet is located at $r=1$)
\bequ
     r\subscr{min} = 0.25, \hspace{1cm} r\subscr{max} = 4.0,
\eequ
\bequ
     \varphi\subscr{min} = 0.0, \hspace{1cm} \varphi\subscr{max} = 2 \pi.
\eequ
For testing purposes the radial range was also reduced in some computations
as presented below.
The outer radial boundary is closed to any mass flow $v(r\subscr{max})=0$,
while at the inner boundary mass outflow is allowed, emulating accretion
onto the central star. However, no matter may flow into the
computational domain from $r\subscr{min}$.
At the inner and outer boundary the angular velocity is set to the value
of the unperturbed Keplerian disc.
To obtain the influence of the planet on the disc accurately, 
a complete ring has to be considered, i.e.
the calculations cover the whole angular range of $2 \pi$.
The values of the physical quantities at $\varphi = 2 \pi$ must agree with
those at $\varphi = 0$, which yields periodic boundary conditions in the
azimuthal direction.
This is implemented in the code by introducing an identification (overlap)
of the first angular gridcell ($j = 1$) with then last one ($j = N_\varphi$).
Care has to be taken that the periodicity conditions are also
fulfilled when solving the matrix equation for the viscosity.

Initially, the matter in the domain is distributed axially symmetric with
a radial profile $\Sigma \propto r^{-1/2}$ which follows in case
of an equilibrium Keplerian disc having a constant viscosity,
and where the inner and outer radius are non permeable.
The initial unperturbed density is normalized such that at
the planet's location (r=1) the dimensionless value of $\Sigma$
is set to 1.236. 
The planet with a mass fraction $q=10^{-3}$ is located initially 
at $r=1$ and $\varphi=\pi$, to avoid any possible problems at the
grid interface. The constant viscosity is set to $\nu = 10^{-5}$.
The temperature profile is given by $T(r) \propto r^{-1}$ which follows
from the specified disc thickness $h_r = 0.05$. This temperature
distribution remains fixed throughout the calculation. 
The radial velocity $v$ is set to zero, and the angular velocity $\omega$ is
set to the Keplerian value of the unperturbed disc,
taking into account the rotation of the coordinates.
Around the planet we then introduce an initial density reduction whose
approximate extension is obtained from the magnitude of the viscosity
(Artymowicz \& Lubow 1994).
This initial lowering of the density is assumed to be axisymmetric; the
radial profile $\Sigma(r)$ of the initial setup is displayed in Fig.~3
below.
The starting model is then evolved in time and the accretion rate onto
the planet is monitored. The amount of matter which is allowed to accrete 
onto the planet is specified in the next section. 

\subsubsection{A simple test of the viscosity treatment}
In general the advective and force terms of the
code have been tested on problems such as shock tubes,
Sedov explosions and other viscous problems (Kley 1989).
For this special problem we found it worthwhile to perform an additional
test of the viscous terms, which had to be newly implemented in the
given ($r-\varphi$) geometry.
We chose the axisymmetric problem of an expanding viscous ring
where the initial surface density has a $\delta$-function profile, which
spreads under the influence of a small viscosity and negligible pressure.
Here the analytical solution for the surface density of
the ring $\Sigma_r$ (with an assumed Keplerian rotation)
can be given in terms of Bessel functions (eg. Pringle 1981)
\[   \label{ring} 
    \Sigma_r(r,t) = \frac{C}{\tau x^{1/4}} 
      \exp{\left(-\frac{1 + x^2}{\tau}\right)}
           I_{1/4} \left(\frac{2 x}{\tau}\right),
\]
where $x$ denotes the normalized radius $x=r/r_0$, with the initial ring
located at $r_0$, $C=1/(\pi r_0^2)$ and $\tau$ denotes the dimensionless
time ($\tau=t/t_v$) in units of the viscous spreading time
$t_v = r_0^2/(12 \nu)$.
Since it is not possible to represent an initial $\delta$-profile decently
on a computer, we chose as the initial starting time $\tau = 0.16$
(cf. Flebbe et al. 1994). Then the ring has already spread a little bit.
%  Figure 1
\begin{figure}
\epsfxsize=8.5cm
\epsfbox{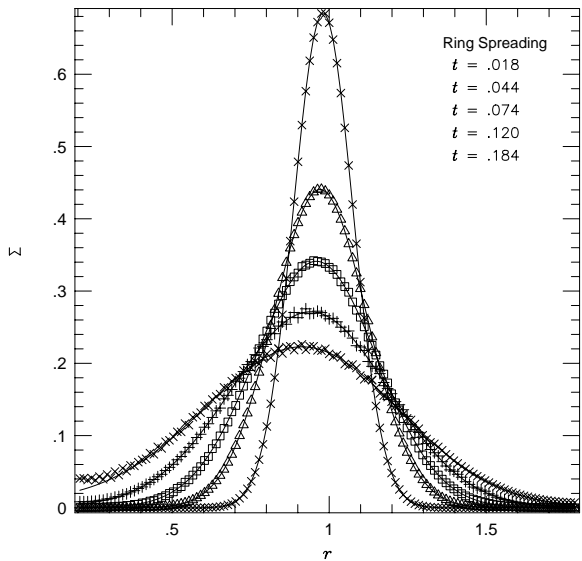}
  \caption{The surface density $\Sigma$ for various times in units of
    the viscous timescale $\tau$. The calculation started with the analytical
     profile at $t = 0.016 \tau$. The solid lines indicate the analytical
     solution described in Section (\ref{ring}).}
\end{figure}
For a constant (dimensionless) viscosity $\nu = 10^{-6}$ and an initial
location of the ring at $r=1$,
the results of a test calculation are displayed in Fig.~1. 
The time evolution follows the analytical results (solid lines)
very well, only at late
times, after the spreading wave is reflected at the inner boundary (which
was closed here), the deviation becomes stronger. Here 128 radial gridcells
were used, and only 5 in the azimuthal direction, because of the
axial symmetry of the problem.
The full equations (forces, and all viscosity terms) were
integrated, for a very low temperature.
The result shows that even in case of very small viscosities the evolution
can be obtained accurately with the present numerical scheme.
No artificial viscosity has been used here,
which causes the oscillations at the inner boundary at late times.
Note, that about 140,000 timesteps were required for the whole evolution.

The test was repeated with $128\times 128$ gridcells and a viscosity
of $10^{-5}$ with an initial random
density perturbation of $1 \%$, in a rotating
coordinate frame using an artificial viscosity ($C_a=0.5$). 
This calculation has parameters which are very similar to those used
in the planet calculations, only the planet mass has been set to zero.
The evolution again agrees well with the analytical
results, and remains axisymmetric.
This, although no non-axisymmetric problems have been tested
these results indicate the accuracy of the numerical scheme in following
the long term evolution of viscously evolving systems.

These tests imply that the numerical viscosity of the scheme lies
below $\nu = 10^{-6}$. One may argue that, as this test describes
only an axisymmetric situation, it does not refer to the case of
an embedded planet with a very high density contrast between
disc and gap region. However, as will be shown below (Tables 3 and 4),
we ran disc models using a physical viscosity as low as $\nu = 10^{-6}$
and even a zero viscosity model. Both were run for about 1000
orbital periods of the planet. In the zero viscosity case the accretion rate
onto the planet was reduced by a factor of 100 over the $\nu = 10^{-6}$
case, which is another strong indication that the numerical viscosity
of the scheme lies definitely below $\nu = 10^{-6}$. These
findings are supported in general by the calculations of Bryden et al. (1998).

\section{Accretion through the Gap}
To study the influence of various physical parameters on the accretion
process in the gap region we construct a reference model with a specific
physical condition. Then some conditions are varied and their influence
on the solution is studied.
The basic model is set up according to section (\ref{bounds}), and is then
evolved in time until a quasi stationary state has been reached. 
Typically, the runs cover about $400$ orbital periods of the planet.

\subsection{Modeling Accretion onto the planet}
The mass accretion onto the planet is achieved by a reduction of the
mass density inside the Roche lobe of the planet.
Basically, the density is reduced by a factor of
$1 - f\subscr{red} \Delta t$ per time
step, where $\Delta t$ is the actual magnitude of the time step.
The rate at which mass is taken out is twice as high in the inner half
of the Roche lobe than in the outer part.
On average, for the standard value $f\subscr{red}= 1/2$,
this translates into a half emptying time of the Roche lobe
of about $t_{1/2} = 2 \log 2$ which is approximately $1/4$ orbital periods
of the planet.
The mass taken out is monitored and is assumed to have been accreted
by the planet. It is not added to the planet's mass though, i.e. $q$ remains
constant.
 
\subsection{The standard model}
Our reference model has the basic parameters as given in Table~1.
The angular velocity of the rotating frame is $\Omega = \Omega_p$
(\ref{t_0}) which is one in the dimensionless units. 
%  Table 1
\begin{table}
\caption[]{Parameter of the standard model, in dimensionless units}
\begin{tabular}{ll}
$r\subscr{min}$, $r\subscr{max}$   &  $0.25, \, \, 4.00$ \\
Grid resolution   &     $128 \times 128$  \\
Mass ratio planet/star   &   $q = 10^{-3}$ \\
Constant viscosity   &   $\nu = 10^{-5}$ \\
Artificial viscosity   &   $C\subscr{art} = 0.5$ \\
Vertical disc height  & $H/r = 0.05$ \\
Rotating frame   &  $\Omega = 1.00$ \\
Mass accretion   &  $f\subscr{red} = 0.50$ \\
\end{tabular}
\end{table}
%  Figure 2
\begin{figure*}
\epsfxsize=\textwidth
\epsfbox{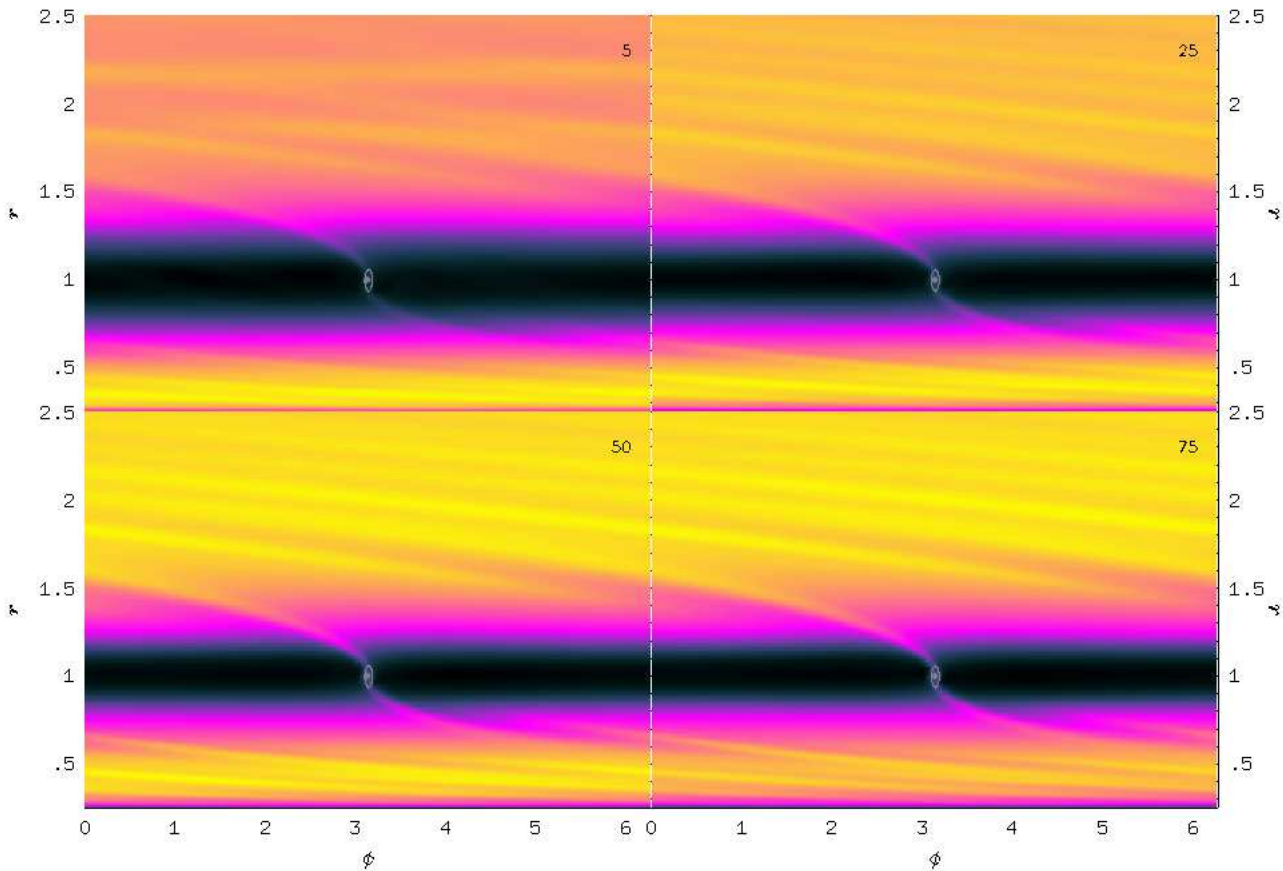}
  \caption{Grayscale plot of the surface density of the standard model
    at four different times,
    given in units of the orbital period of the planet.
    The scaling has been chosen such that black refers to $10^{-4}$
    and white to $1.0$. The size of the planet's Roche-lobe is indicated
    by the solid line around ($r=1, \varphi=\pi$)}
\end{figure*}
The planet's influence on the disc manifests itself in the creation of
spiral disturbances in the surface density. The spiral pattern emerges
very early during the evolution, already within the first 5 orbital
periods, and the settles fast into a stationary state.
The density is displayed
in a gray scale plot in Fig.~2 at four different times. The final
pattern consists of two tightly wound spirals on either side of the planet.
On both sides, a {\it primary} spiral starts at the location of
the planet. The secondary spirals start near the
$L_4$ and $L_5$ points at $\varphi \approx \pi \pm \pi/3$
outside and inside of the planet.
The tightness and 
the existence of additional spirals to the primary ones depend mainly
on the equation of state, basically on the temperature in the disc
(see below). Note, that in the corotating frame ($\Omega =\Omega_p$)
the spirals are stationary and do not move through the grid.
The density decrease (mass loss) inside of the planet is caused by
the open inner boundary condition, where mass may leave the system
to be accreted by the star. 

The initial mass distribution approximates a disc already in a perturbed
state including a gap.
%  Figure 3
\begin{figure}
\epsfxsize=8.5cm
\epsfbox{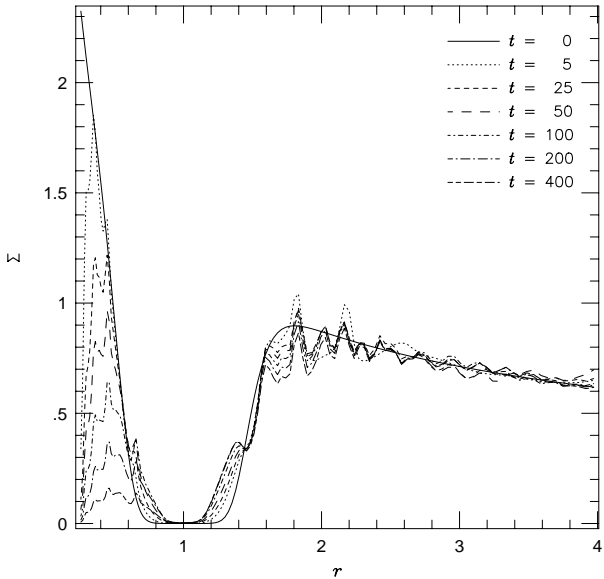}
  \caption{The Surface density $\Sigma(r)$ of the standard model
     opposite of the planet
     at eight different times.}
\end{figure}
The gap also evolves slowly during the evolution and begins to narrow
in the central parts
somewhat from its initial radial extension. This is more clearly visible
in Fig.~3, where $\Sigma (r)$ is plotted at a fixed angle $\varphi=0$
in opposition to the planet at different times. 
The initial axially symmetric profile is given by the solid line.
Clearly visible are the different peaks produced by the spiral waves.
The location and shape of the spirals are determined very early during the
the evolution on dynamical timescales ($\Omega_p^{-1}$). The lowering of
the density ($\Sigma$) in the outer parts is due to the accretion onto
the protoplanet.
While in the inner part ($r<1$) it is primarily due to mass loss through
the open inner boundary.

%  Figure 4
\begin{figure}
\epsfxsize=8.5cm
\epsfbox{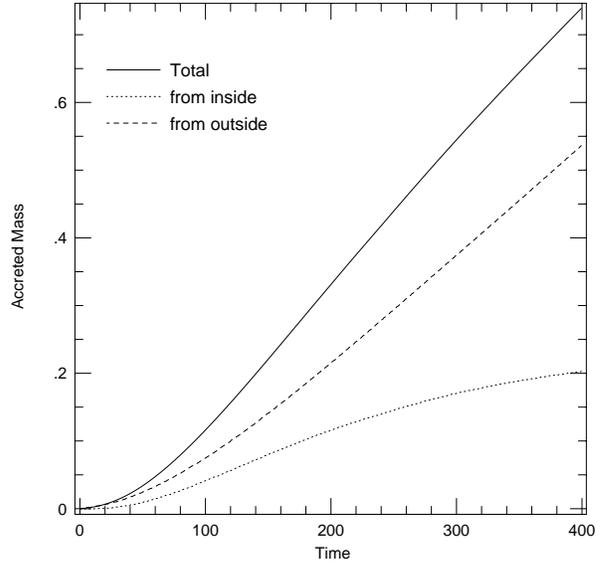}
  \caption{Accreted mass onto the planet in dimensionless units, from
     the inside and outside of the planet for the standard model.}
\end{figure}
The mass accretion of the planet is plotted in Fig.~4, where it is
distinguished whether the mass came from the outer parts ($r > 1$)
of the disc or the inner parts ($r < 1$), respectively.
In the computations we monitored continually the matter content in the
disc regions exterior and interior of the planet's orbital
distance to the star. The mass accretion rate from
the outer region approaches approximately a constant
value after $t \approx 200$ (linear part of the dashed curve in Fig.~4),
when the streams are fully developed.
Initially the accretion rate is very small because of the pronounced imposed
gap.
The accretion from the inner region begins to diminish because of the loss
of mass through the inner boundary.
The actual physical value of the mass accretion rate depends on the assumed
total mass initially in the disc, which is here in dimensionless
units $M(0) = 33.0$.
Thus about $2.24\%$ of the total initial mass was accreted during
400 orbital periods of the planet.

From Fig.~5 we infer that indeed the mass accretion rate
{\it from the outside}
approaches a constant value, $\Mdot_P = 1.63 \, 10^{-3}$ in dimensionless
units, where the unit of time is $P_p$ (Eq.~\ref{Pp}). Assuming that the
initial total mass in the disc (from $r\subscr{min}$ to $r\subscr{max}$) is
$10^{-2} \Msol$ and that the planet orbits the star at a distance of $5.2$AU,
then the unit of the mass flow rate as given in Fig.~5
(and subsequent figures)
refers to $2.67 \, 10^{-2} M\subscr{Jup}/yr$. Hence, for our
{\it standard model} we obtain after the initial transient the constant value
\bequ  \Mdot\subscr{P}(\mbox{out}) = 4.35 \, 10^{-5} M\subscr{Jup}/yr  
  \label{macc-2q}
\eequ
for the equilibrium mass accretion rate from the outer disc onto
a protoplanet with a mass ratio $q=10^{-3}$ and the disc parameter
$\nu = 10^{-5}$ and $H/r = 0.05$.
%  Figure 5
\begin{figure}
\epsfxsize=8.5cm
\epsfbox{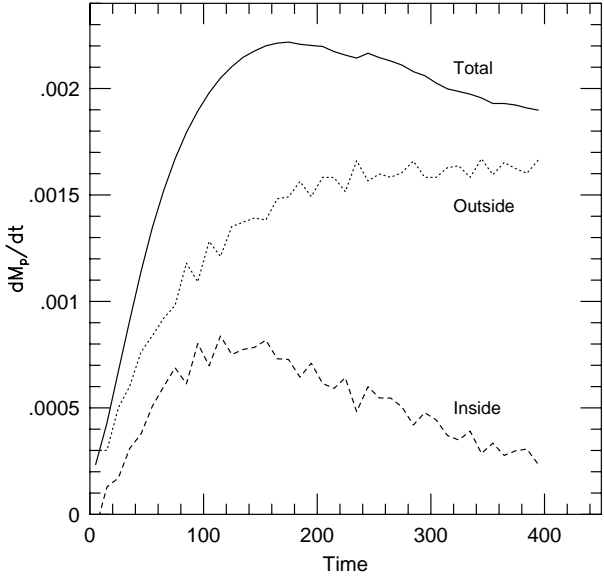}
  \caption{Mass accretion rate onto the planet in dimensionless units, from
     the inside and outside of the planet for the standard model.}
\end{figure}
\subsection{Influence of Numerical Parameters}
The computation of the dynamics of the tenuous streams through the
gap may depend on numerical effects such as artificial viscosity, resolution,
boundary condition and so on. Thus, to obtain reliable estimates
of the magnitude of a possible accretion rate through the gap, we kept the
physical conditions of the model unchanged and varied only numerical
input parameters.
After these tests on the sensitivity of the results on the numerics
we may concentrate later on the influence of the physical conditions
on the gap dynamics.
%  Figure 6
\begin{figure}
\epsfxsize=8.5cm
\epsfbox{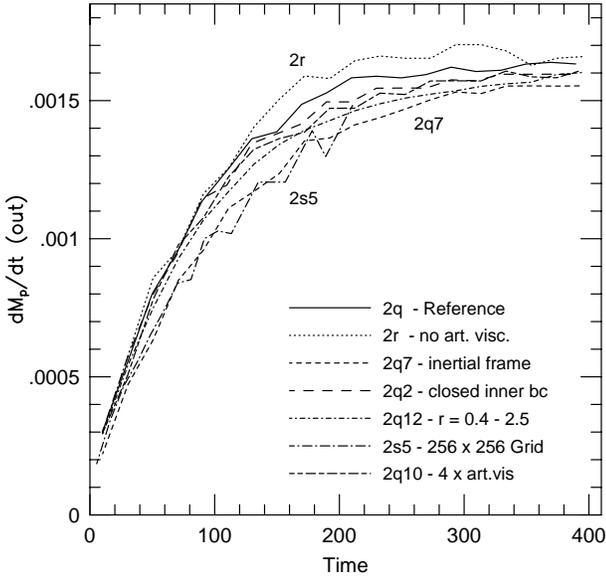}
  \caption{Mass accretion rate from outside of the planet for
     different test cases with varying parameters (Table 2). The solid
     curve denotes the standard model (2q). The variations from
     the standard model are given by the labels.}
\end{figure}
%  Table 2
\begin{table}
\caption[]{Parameter variations of from the standard model (Table 1) for
the models displayed in figures 6, 7, 9, and 11}
\begin{tabular}{lll}
Name  & Parameter  & Description \\
\hline
 2q   &  Standard   &  Reference Model \\
 2q2  &  $v(r\subscr{min}) =0$  &  Closed inner boundary \\
 2q7  &  $\Omega = 0$  &  Inertial frame \\
 2q12 &  $r\subscr{min} =.4, r\subscr{max}=2.5$ & Smaller physical domain \\
 2s5 &  $256 \times 256$  & Higher resolution \\
\hline
\multicolumn{3}{c}{Variations of Artificial Viscosity (Fig6., Fig. 7)}\\
 2q10 &  $C_{art} =2.0$ & Four times larger bulk\\
      &                 & artificial viscosity \\
 2n &  $C_{art} =2.5$ & Large Shear Art. Vis. \\
 2l &  $C_{art} =0.5$ & Std. Shear Art. Vis. \\
 2p &  $C_{art} =0.1$ & Low Shear Art. Vis. \\
 2r  &  $C_{art} =0.0$ & No Artificial Viscosity\\ 
\hline
\multicolumn{3}{c}{Variations of Initial gap size (Fig. 9)}\\
 2q1   &             &  No initial gap \\
 2q19  &             &  50\% smaller gap \\
 2q20  &             &  50\% larger gap \\
\hline
\multicolumn{3}{c}{Variations of accretion modeling (Fig. 11)}\\
 2q3   &  $f\subscr{red}= 1/20$  &  Reduced accretion rate \\
 2q6   &  $f\subscr{red}= 5/2$   &  Enhanced accretion rate \\
 2t5   &  $f\subscr{red}= 1/2$   &  1/2 accretion radius \\
 2t6   &  $f\subscr{red}= 5/2$   &  1/2 accretion radius \\
\end{tabular}
\end{table}
In Fig.~6 the main results of our studies are displayed.
The single variations from the standard models are named in the figure,
and are listed in Table~2, all other parameter are identical to the
reference model (2q) which is
given by the solid line.
Some of the irregularity in the results is caused by taking the derivative
of data of the mass of the planet.
Three models with a somewhat larger deviation (2r, dotted;
2s5 long dash-dotted; and 2q7 short-dashed)
are explicitly labeled.
The model with the vanishing artificial viscosity (2r) has a slightly 
higher accretion rate. The model with the larger resolution (2s5, only run to
$t=200$) lies slightly below. Both of these models are noisier
than the others due a too small artificial viscosity.
The noise in $\Mdot$ is usually larger for models with higher resolution
because for smaller gridcells numerical effects (grid to grid
oscillations) tend to increase. However, the artificial viscosity
coefficient ($C_{art}$) has not been increased for model (2s5).
The model that was run in the inertial frame (2q7) has possibly more
numerical diffusion in the whole computational domain, as the planet
and the spiral wave pattern are not stationary but move through the
grid. Nevertheless, the agreement with the standard model is very good.
%  Figure 7
\begin{figure}
\epsfxsize=8.5cm
\epsfbox{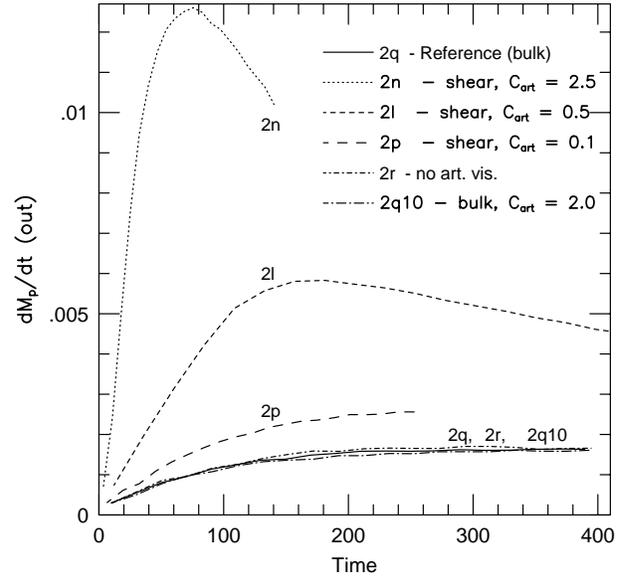}
  \caption{Mass accretion rate from outside of the planet for
     different types of artificial viscosity.}
\end{figure}

All other models lie very closely to the standard model.
The model with the increased artificial viscosity (2q10, long-short dashed) 
and the model with the closed inner boundary (2q2, long dashed) give nearly
identical results.
The model with the smaller computational domain (2q12, short dashed-dotted)
tests on one hand
the influence of a variation the location of the radial boundaries
(the reference model has $r\subscr{min}=0.25$ and $r\subscr{max}=4.0$)
and on the other hand the influence of an
increased resolution in the radial
direction, since the same amount of radial grid points (128) was used. 
Both of these effects have basically no influence on the results.
In summary, all models despite their parameter variations are consistent
with the standard model.

We also would like to point out that the open inner
boundary does not affect the accretion rate from the outside of disc.
This indicates an independent mass flow from the inner and outer regions
onto the planet with no or little interference where the inner
mass accretion rate
appears to be a given fraction of the outer one.
In the computations we monitored continually the matter content in the
disc regions exterior ($r>a$) and interior ($r<a$) of the planet's orbital
distance to the star. From the model with the
closed inner boundary (2q2) we obtain that the mass accretion rate
from the {\it inner}
region $\Mdot_p(in)$ also settles to a constant value with
\bequ  \label{mdotin}
\Mdot_p(in)/\Mdot_p(out) = 3/5.
\eequ
We concentrated here
on the accretion from the outside since the mass on the inner side may be
depleted during the evolution of the whole disc (for example
by accretion onto the star or by an inner planet).
These conditions were simulated by the open inner boundary.
On the contrary, we may argue that the mass flow rate from the outer region
alone gives a lower limit of the total accretion rate onto the system,
for the given physical conditions.

Hence, we may conclude that we can detect a
possible mass accretion rate onto a planet through a gap 
quite accurately, independent of numerical issues. 
And that a relatively moderate grid size of $128 \times 128$
seems to be sufficient for
an estimation of the mass accretion rate. However, the details of the
flow in the vicinity of the planet need to be computed with
much higher resolution. 
The time needed for reaching an equilibrium mass flow rate depends
on the viscosity and is approximately $400$ orbits for $\nu = 10^{-5}$.
%  Figure 8
\begin{figure*}
\epsfxsize=\textwidth
\epsfbox{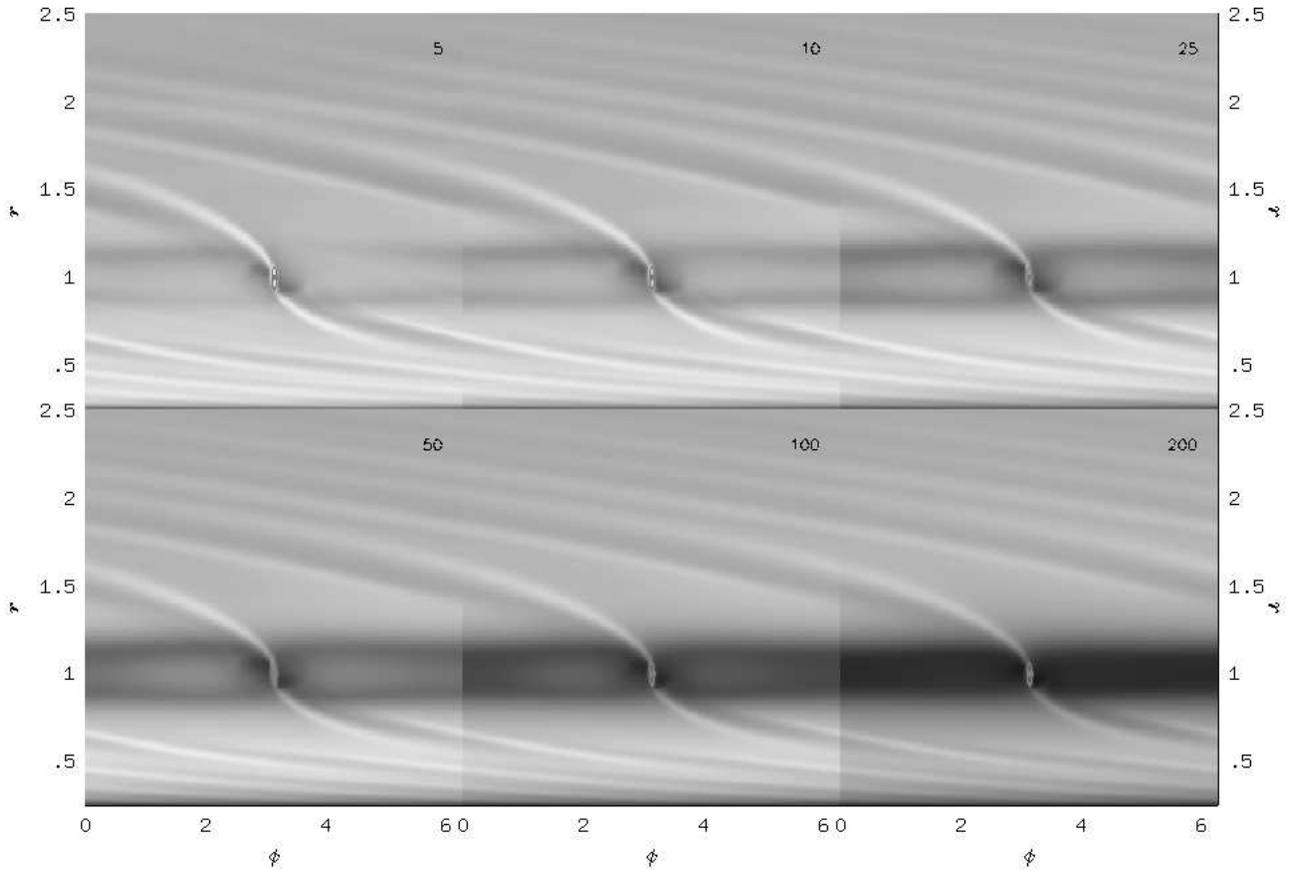}
  \caption{Grayscale of the surface density of the model
    with no initial gap at six different times.}
\end{figure*}
\subsubsection{Remarks on the artificial viscosity}
Since numerical instabilities tend to originate in the low density gap region,
there is, as has been seen already in the preceding paragraph,
the necessity of using an artificial viscosity. As mentioned above
we found it necessary that this additional viscosity is only applied to
the bulk viscosity part, and not the shear viscosity components.
To illustrate the reasoning, we ran some models with a shear artificial
viscosity having different magnitudes of the coefficient $C\subscr{art}$.
To distinguish this effect, it is not very useful to look at the
two-dimensional flow fields since they are always very similar. But the
crucial mass accretion rate tends to have a strong sensitivity to this
purely numerical issue.

From Fig.~7, where the mass accretion rate
for different models with a varying type of artificial viscosity is plotted,
it is apparent that in the models using an artificial viscosity which operates
also on the shear contributions of the viscous stress tensor, the mass
accretion rate depends strongly on the magnitude of the viscosity coefficient
$C\subscr{art}$. The model having the same $C\subscr{art}$ as the reference
model (2l, short dashed line) has about a three times larger $\Mdot_p$ rate.
Additionly, the mass accretion rate depends strongly on the magnitude
of the coefficient. An increase/decrease in $C\subscr{art}$ leads subsequently
to an increase/decrease in the accretion rate $\Mdot_p$.
The model (2n, dotted) has a five times higher $C_{art}$, resulting in such
a large fictitious mass accretion onto the planet that the mass reservoir
outside of the planet begins to deplete.
Only for very low
values of $C\subscr{art}$ (2p, long dashed) the curves seems to
approach the zero artificial
viscosity value (2r, short dashed dotted).
Apparently the shear artificial viscosity acts as an additional physical
viscosity influencing strongly the estimated mass accretion rate.

In case of an artificial viscosity only operating on the bulk
parts of the viscous stress tensor (2q10, long dashed dotted)
there is no apparent dependence of
the accretion rate on $C\subscr{art}$; the curve with
$C\subscr{art}=2.0$ which is four times standard, is nearly identical
to the zero artificial viscosity case.
One may argue here, that if runs with no artificial viscosity are possible
why bother at all. The reason is that for higher resolution, or for
different physical viscosities the usage of an artificial viscosity may be
warranted to prevent numerical instabilities. 
Thus, we conclude that the bulk artificial viscosity is the selection
of choice, since firstly, it does not change important physical
effects, and secondly it is sufficient to keep the numerical scheme stable.
From these tests we may also infer that the mass accretion depends crucially
on the type of viscosity used, and will be reduced/increased
for smaller/larger physical viscosity coefficients (see below).
\subsubsection{Varying the initial condition}
The results displayed so far indicate an asymptotic, constant value
for the mass accretion rate (Eq. \ref{macc-2q}) given a specified 
value of the initial mass in the region outside the planet.
As these results were obtained with an initial gap around the planet,
it is interesting to study the situation with no such initially
imposed gap. Hence, we constructed models with varying extensions
of the initial gap. One model (2q1) had initially no gap
at all but rather the equilibrium surface density distribution
$\Sigma (r) \propto r^{-1/2}$ throughout the whole
computational domain. Two additional models with a $50\%$ smaller
and larger radial initial extension of the gap have been calculated
(see Table~2).

In the case with no initial gap at all the
torques generated by the planet begin to open
the gap from 'scratch' and at $t =5$ the spiral wave pattern and the general
shape of the gap are already clearly visible (Fig.~8). In the following 
evolution the gap clearing proceeds rapidly where
the regions near the Lagrangian points $L_4$ and $L_5$ which
are located at $\Delta \varphi = \pm \pi/3$ from the planet have the longest
clearing time.
The final shape of the gap and spiral pattern (eg. at $t=200$)
is nearly indistinguishable from the reference model with an initial gap.
The mass accretion rate is of course initially very high for this model
but for later times is approaching the accretion
rate of the previous standard model (Fig.~9).
The $\Mdot_p$ curves of the models with the 50\% smaller/larger initial gap
size also approach the standard curve from above/below (Fig.~9). 
Hence, the models seem to converge to the same final equilibrium accretion
mass accretion rate $\Mdot_p$ despite of the very different initial conditions.
 
The process of the gap clearing for the model with no initially imposed
gap is exemplified
in Fig.~10, which should be compared to the standard model
in Fig.~3 above.
Initially the gap begins to clear fastest at radii slightly larger
and smaller than the planet, i.e. at its boundaries where the
tidal torques are highest.
Only at very late times the density in the central
region of the gap reaches the plateau present in the standard model
from the very beginning.
Clearly visible is again the stationarity of the location of the
spirals which appear at specific radii after very few orbits
and remain stationary there throughout the evolution.

From the test models, with varying initial gap sizes it is
clear that evolution takes place on a viscous time scale, which
is several thousand orbits for the constant viscosity $\nu = 10^{-5}$.
%  Figure 9
\begin{figure}
\epsfxsize=8.5cm
\epsfbox{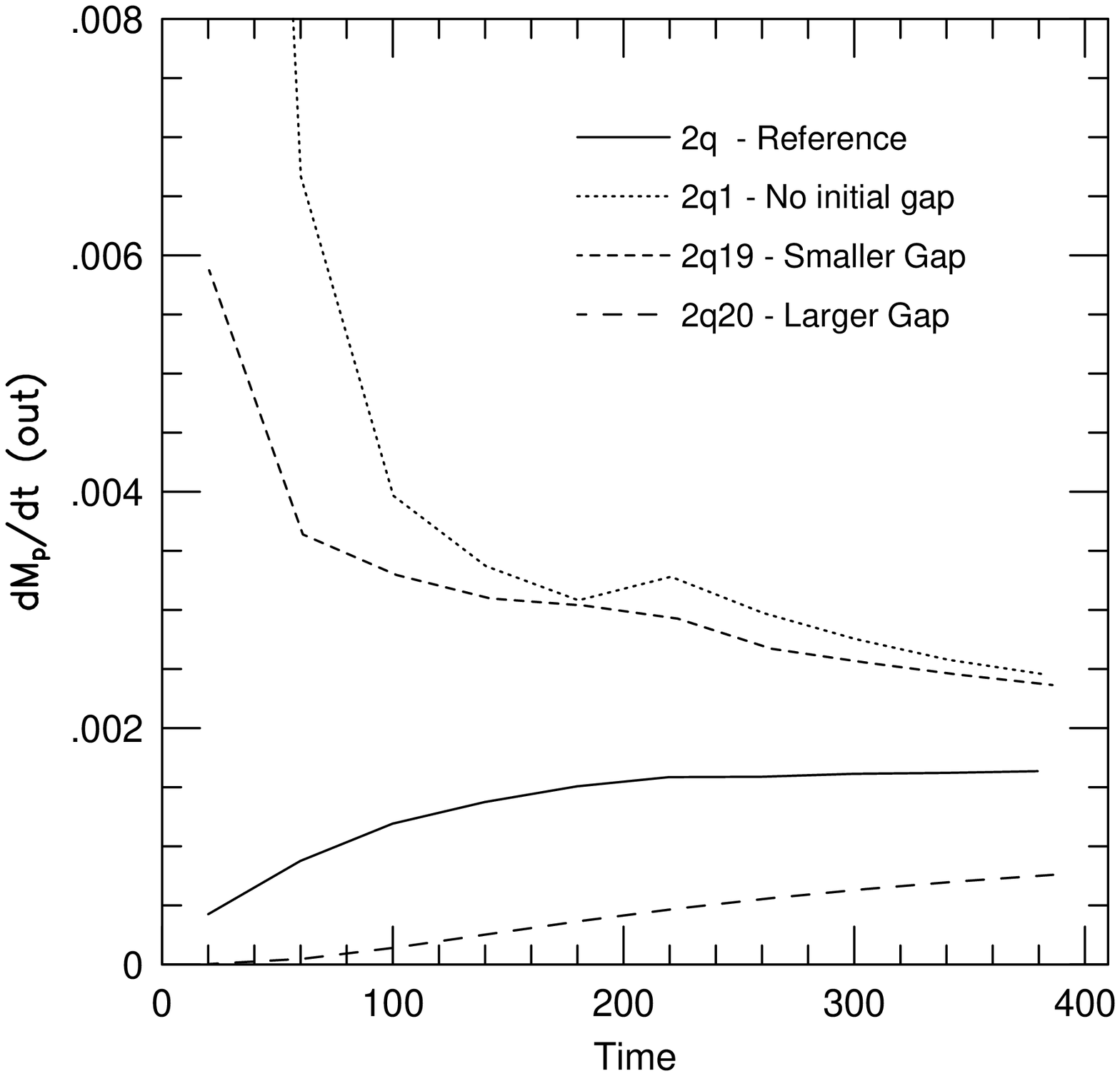}
  \caption{Mass accretion rate from outside of the planet for
     the standard model and models with different gap sizes.}
\end{figure}
%  Figure 10
\begin{figure}
\epsfxsize=8.5cm
\epsfbox{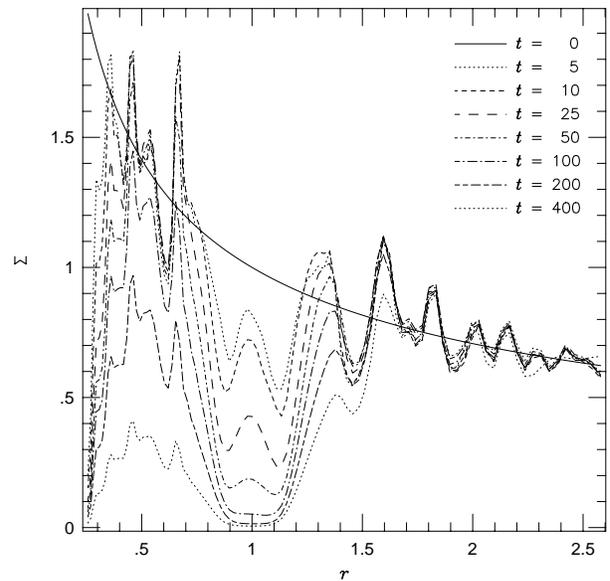}
  \caption{The Surface density $\Sigma(r)$ opposite of the planet
     (at $\varphi=0$) for the model with no initial gap (2q1).
     The solid line indicates the initial
    density distribution.}
\end{figure}
\subsubsection{Influence of the accretion radius inside the Roche lobe}
%
%  Figure 11
\begin{figure}
\epsfxsize=8.5cm
\epsfbox{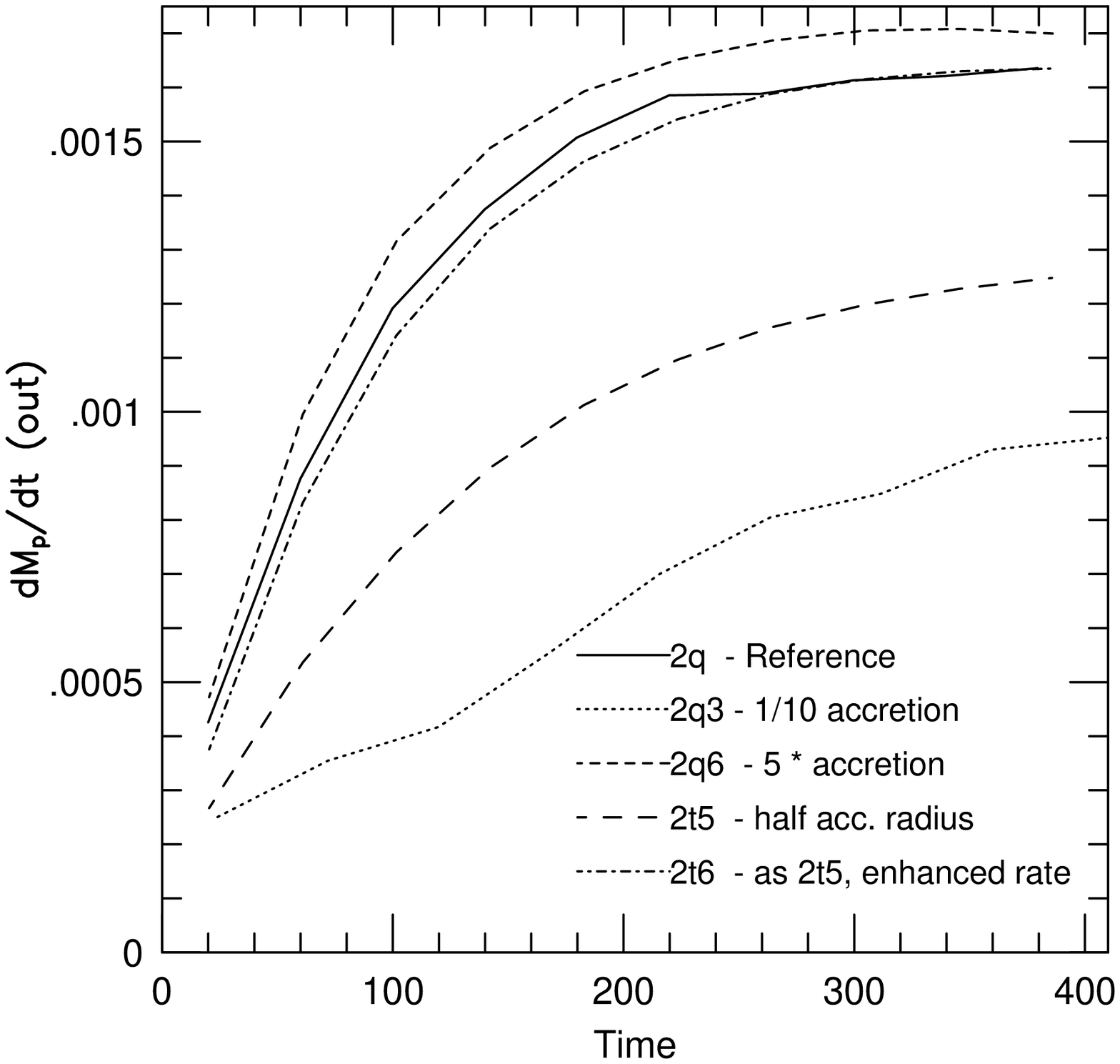}
  \caption{Mass accretion rate for different accretion radii within
     the Roche lobe.}
\end{figure}
In section 3.1 we explained how the accretion process onto the planet is
modeled numerically. The value of the reduction
factor $f\subscr{red}$ may play an important role in determining the
accretion rate onto the planet. To check its influence we ran
models where $f\subscr{red}$ has been varied (see Table~2).
The results (Fig.~11) indicate
that firstly a larger increase (5 times bigger in model 2q6)
does not alter the achieved maximum accretion rate significantly,
while a ten times smaller factor (2q3)
reduces the final accretion rate only by about $40\%$.
In all the calculations presented so far the accretion radius (i.e. the domain
in which mass was taken out to simulate accretion) was the whole size of the
Roche lobe, with an enhanced accretion within the inner
half of the Roche lobe. Two additional models where the size of the
accretion region was reduced to only the inner half of the lobe with an
enhancement in the inner quarter were run in the inertial frame. In the first
model (2t5) the reduction factor was kept (despite the smaller
accretion radius) at $f\subscr{red}=0.5$ while in the second model (2t6) it was
increased by a factor of 5. While the first model (2t5) has about a $30\%$
smaller accretion rate, the latter model reaches the same value as the
standard model.

We may conclude that independent of the details of the modeling
of the accretion process within the planet's Roche lobe,
there exists a well defined {\it maximum} accretion rate as given
by our standard model (Eq.~\ref{macc-2q}). 
If the planet does not fill its Roche lobe entirely this maximum rate
may be reached. For comparison, Jupiter's radius is only $1.34 \, 10^{-3}$
of its Roche radius.
This maximum achievable accretion rate may be compared to the equilibrium
rate of a stationary accretion disc which is given by
\bequ  \label{mdotacc} 
     \Mdot\subscr{acc} = 3 \pi \nu \Sigma.
\eequ
In dimensionless units we obtain $\Mdot\subscr{acc} = 5.9 \, 10^{-4}$ 
(for $\Sigma = 1.$ and $\nu=10^{-5}$) which is
about 2.8 times smaller than our maximum accretion rate.
At a first glance this may be surprising, but one has to remember that
(\ref{mdotacc}) refers to an undisturbed disc. Boundary effects in an
accretion disc
(for example near the central star) may increase the equilibrium
accretion rate
considerably (eg. Frank, King \& Raine 1992). The presence of a planet
in the disc perturbes the disc flow and acts as a inner boundary
to the outer parts of the disc at $r>a$, modifying the
relation (\ref{mdotacc}). Note, with boundary effect we do not refer
to the conditions at the inner boundary ($r\subscr{min}$) of the computational
domain (closed or open) which do not
have any influence on the accretion rate from the outside as shown above.
Additionally, an increased radial pressure gradient in the gap region
will increase the accretion rate (see Sect. \ref{subsec-height} below).
Both, the effect of the planet on the outer disc,
and pressure effects may account
for a possible slight enhancement of $\Mdot_p$ above $\Mdot\subscr{acc}$.

Having tested the effects of several numerical aspects of the code and
boundary and initial conditions we come now the influence of the
physical properties of the disc.
%  Table 3
\begin{table}
\caption[]{Variations of the physical viscosity and vertical thickness for
the models displayed in Figures 12 and 13, and Table 4.
For the reference model see Table 1}
\begin{tabular}{lll}
Name  & Parameter  & Description \\
\hline
\multicolumn{3}{c}{Variations of the viscosity (Fig. 12)}\\
 2q   &  $\nu=1 \, 10^{-5}$  &  Reference Model ($H/r=0.05$)\\
 2q4  &  $\nu=4 \, 10^{-5}$  &  Four times larger visc.\\
 2q14  &  $\nu=2 \, 10^{-5}$  &  Two times larger visc.\\
 2q15  &  $\nu=5 \, 10^{-6}$  &  Two times lower visc.\\
 2q5  &  $\nu=2.5 \, 10^{-6}$  &  Four times lower visc.\\
 2q23  &  $\nu=1 \, 10^{-6}$  &  Ten times lower visc.\\
 2y1   &  $\nu= 0$  &  Zero viscosity\\
\hline
 2x1   &  $\alpha = 0.004$  &  $\alpha$-type viscosity\\
 2x   &  $\alpha = 0.001$  &  $\alpha$-type viscosity\\
\hline
\multicolumn{3}{c}{Variations of the disc height (Fig. 13)}\\
 2qz  &  $H/r=0.1$  &  Two times larger $H/r$\\
 2q8  &  $H/r=0.075$  &  1.5 times larger $H/r$\\
 2q17 &  $H/r=0.025$  &  2 times lower $H/r$\\
\end{tabular}
\end{table}
\subsection{Influence of the physical viscosity}
\label{subsec-physvis}
The calculations presented so far have been performed with the
constant kinematic viscosity $\nu = 10^{-5}$. 
It is to be expected that the mass accretion rate depends on the
magnitude of the physical viscosity. Thus, we performed different
calculations where the kinematic viscosity coefficient $\nu$ was
varied from its standard value $10^{-5}$. For comparative purposes we
also used an $\alpha$-type viscosity with two different values of
$\alpha$.
For all the models the remaining physical parameters, in particular the
scale height (temperature), are identical to the standard model (2q).
Note also, that the initial density profile and the gap size
was identical for all models. The model parameter are given in Table~3,
The main results are displayed in Fig.~12. Note, that the two models
with the lowest (2q23) or even zero (2y1) viscosity were 
evolved for about 1000 orbits. They are not included
in the figure since their accretion rates 
which are stated in Table~4 are very small.
%  Figure 12
\begin{figure}
\epsfxsize=8.5cm
\epsfbox{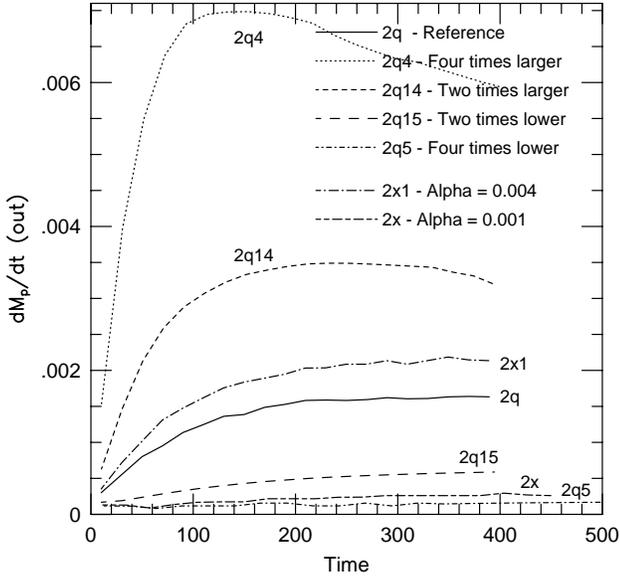}
  \caption{Mass accretion rate from the outside of the planet for
     different physical viscosities.}
\end{figure}
Obviously the mass accretion rate depends strongly on the magnitude
of the viscosity. The decrease in $\Mdot_p$ for the larger viscosity models
(2q4, 2q14) is a result of the mass depletion of the outer region of the disc.

From the results we construct Table~4, where the accretion
rates are listed for different
values of the viscosity (both in dimensionless
units, where $[\dot{M}_p] = 2.67 \, 10^{-2} M\subscr{Jup}$/yr).
Also stated is the growth timescale $\tau_g$ of the planet.
The values of the accretion rate for the higher viscosity
models (2q4, 2q14) are
taken at the maximum value before mass depletion sets in.
%  Table 4
\begin{table}
\caption[]{Dependence of the mass accretion rate on the viscosity,
   both given in dimensionless units. In the third column the
   growth time scale $\tau_g = m\subscr{p}/\dot{M}\subscr{p}$ is
   given in years for a one Jupiter mass planet at a distance $r\subscr{J}$
   from the star.}
\begin{tabular}{llll}
  Model  &    Viscosity     &  $\Mdot_p(out)$   &  $\tau_g$  \\
         &    [$10^{-5}$]   &  [$10^{-3}$]      &    [yrs]   \\
\hline
   2y1   &    0      &  0.0003  &   $1.2 \, 10^{8}$  \\
   2q23  &   0.10    &  0.03    &   $1.2 \, 10^{6}$  \\
   2q5   &   0.25    &  0.18    &   $2.1 \, 10^{5}$ \\
   2q15  &   0.50    &  0.64    &   $5.8 \, 10^{4}$  \\
   2q    &   1.00    &  1.63    &   $2.3 \, 10^{4}$  \\
   2q14  &   2.00    &  3.4     &   $1.1 \, 10^{4}$  \\
   2q4   &   4.00    &  7.0     &   $5.4 \, 10^{3}$  \\
\hline
   2x    &  $\alpha=0.001$   &  .28     &   $1.3 \, 10^{5}$  \\
   2x1   &  $\alpha=0.004$   &  2.2     &   $1.7 \, 10^{4}$  \\
\end{tabular}
\end{table}
We infer that the mass accretion rate
for larger viscosity depends approximately linearly on the
viscosity with an offset (zero $\Mdot_p$ at $\nu \approx 1.6 \, 10^{-6}$).
For lower viscosities ($\nu \lsim 5 \, 10^{-6}$) however,
the $\Mdot_p$ rates are dropping much faster
than linear.  For a ten times lower viscosity (2q23, $\nu = 10^{-6}$) the
accretion rate onto the planet is reduced by a factor of about 50.
The model with no physical viscosity (2y1) has an accretion rate which
is again 100 times smaller. This is an additional indication of the low
numerical viscosity in the numerical scheme. 

For the models having a constant $\alpha$-type viscosity
(cf. Eq. \ref{nu_alp}), we find that
the model with $\alpha=0.004$ (2x1) refers approximately to the standard
model with $\nu = 10^{-5}$. Which was expected from the initial setup,
where we chose for $\alpha$ such a value that the two viscosities agree
at the radius $r=1$. 
The ratio of $\Mdot_p$
between $\alpha=0.004$ and $\alpha=0.001$ is similar to the one between
$\nu=1.0 \, 10^{-5}$ and $\nu=0.25 \, 10^{-5}$.
We note, that the accretion rate for the $\alpha = 0.001$ model (2x)
refers to $7.5 \, 10^{-6} M\subscr{Jup}/yr$.

In the last column the growth time scale
$\tau_g = m\subscr{p}/\dot{M}\subscr{p}$ is given in years
which may be compared to the calculations of Bryden et al. (1998).
The accretion time scale has to compared to the viscous evolution time
of the disc which is given for the constant viscosity by
\begin{equation}
     \tau_\nu = 1.9 \nu^{-1} \mbox{yrs}.
\end{equation}
Hence we find that for $\nu = 10^{-6}$ the two time scales
are comparable, while for larger $\nu$ the accretion timescale is smaller
than the viscous evolution time scale (for $H/r =0.05$, and $q=10^{-3}$).
\subsection{Influence of the scale height of the disc}
\label{subsec-height}
Here we study the consequences of a varying thickness (temperature)
of the disc on the mass accretion rate keeping the viscosity coefficient
fixed at $\nu = 10^{-5}$. The model parameter are summarized in Table~3.
%  Figure 13
\begin{figure}
\epsfxsize=8.5cm
\epsfbox{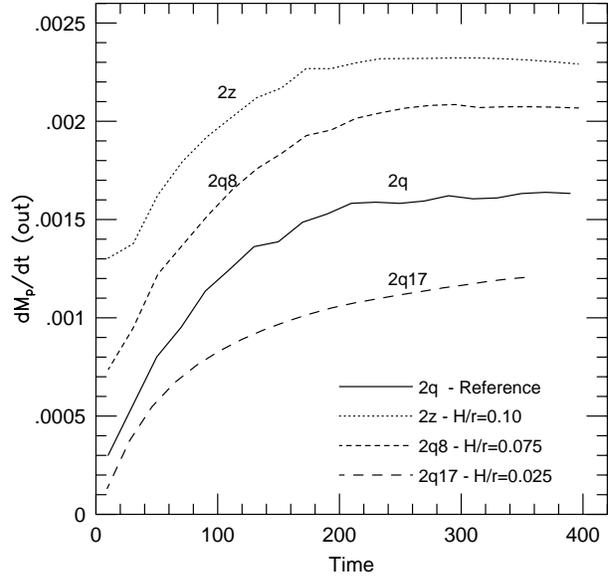}
  \caption{Mass accretion rate from the outside of the planet for
     different values of the disc thickness.}
\end{figure}
From Fig.~13 we infer that (at least for the models
having a smaller $H/r$) the mass accretion rate depends linearly
on the chosen value of $H/r$ with an offset at
$\dot{M}\subscr{p} \approx 8 \, 10^{-4}$.
The variations in the thickness were chosen
with fixed linear spacing which led approximately
to a linear change in the accretion rate.
The model with the largest $H/r=0.1$ (2z) shows a value slightly
too small compared with this relationship.
This deviation from the linear behaviour may be caused firstly
by the mass depletion of
the outer parts and secondly by non-linear effects which begin to set in
for larger internal pressure of the disc. 
For the given grid resolution it is not possible to study models with
a smaller $H/r$ value, since then the radial pressure scale length
would be smaller than the size of one gridcell.

\subsection{Influence of the equation of state}
So far an isothermal equation of state has been used, where the
temperature profile in the disc had a given $\propto r^{-1}$ dependence.
To simulate possible cooling effects and their influence on the viscosity
we performed some additional calculations using a polytropic
equation of state (\ref{poly}). An adiabatic exponent $\gamma=2$ was
used and the constant $K$ has been set to such a value that the
disc thickness is approximately $0.05$ at the radius $1$. In dimensionless
units this refers to $K=0.0025$. The viscosity prescription was also
varied in these models. Starting from the standard constant value
$\nu=10^{-5}$, models with no physical viscosity at all (4y) and with an
$\alpha$-type viscosity were considered.

%  Figure 14
\begin{figure}
\epsfxsize=8.5cm
\epsfbox{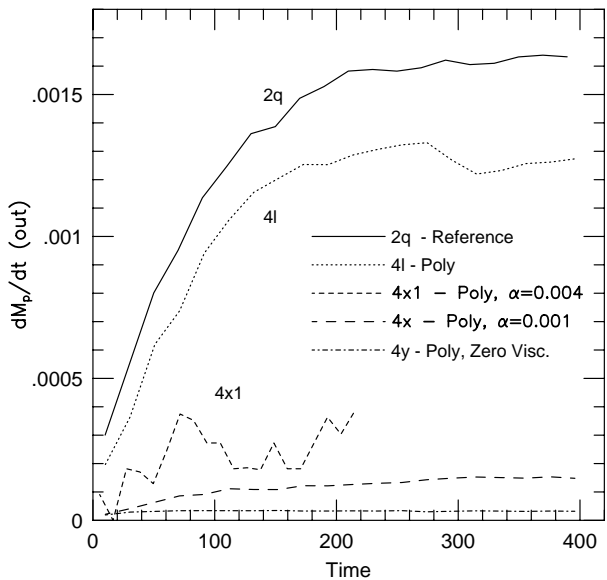}
  \caption{Mass accretion rate from the outside of the planet for
     polytropic models using different parameter.}
\end{figure}
As can be seen from Fig.~14, the model using a constant $\nu$ but a
polytropic equation of state (4l) follows the reference model
closely. The constant $K$ was chosen slightly too small for a better
agreement. Choosing an $\alpha$-type viscosity the mass accretion rate is
also for the larger value $\alpha=0.004$ (model 4x1) reduced from the
constant $\nu$ case (cp. Fig.~12).
A value of $\alpha = 0.001$ (4x) lies also below the isothermal model.
The data for the $\alpha = 0.004$ model are relatively noisy, most likely
because non-linear effects caused by the strong
dependence of pressure and viscosity on the surface density begin to set in.
A detailed analysis of these phenomenon lies outside of the scope
of this paper.

The general trend of an $\alpha$-type viscosity prescription
{\it together} with
a polytropic equation of state is obvious, however.
For polytropic models
the pressure is linked directly to the density,
and because the density and pressure scale height are very small in the gap,
this leads to a reduction of the viscosity for an $\alpha$-type law,
and consequently to a lowering of the mass accretion rate onto the planet.
This does not occur in case of the isothermal equation of state
(see model 2x, 2x1 in Fig.~12),
where the temperature in the gap is not reduced.
Additionally, for an isothermal equation of state the density in the gap
can never vanish, but it can for a polytropic equation of state.
For comparison we calculated also a model with vanishing physical
viscosity (4y). As expected, the mass accretion rate in this
case effectively vanished. Again, a comparison isothermal model with zero
physical viscosity was run for about 1000 orbits which shows
very low accretion onto the planet $\Mdot_p < 3 \, 10^{-7}$ (see Table 4).

\subsection{The size of the gap}
%  Figure 15
\begin{figure}
\epsfxsize=8.5cm
\epsfbox{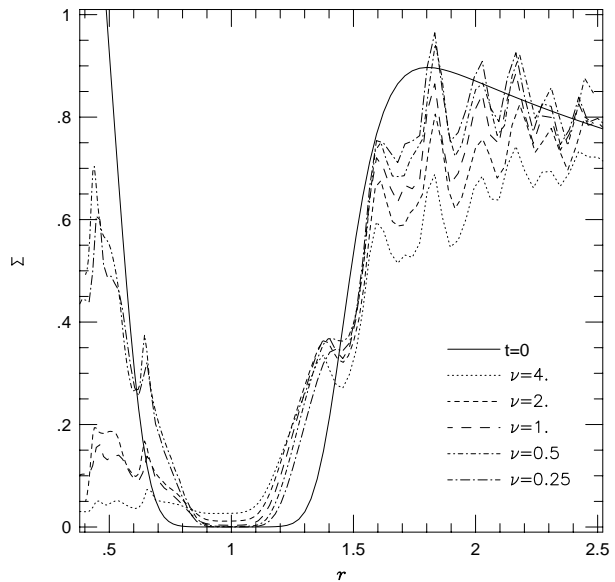}
  \caption{Radial surface density distribution at $\varphi=0$ in
opposition to the planet for various viscosities stated in units
of $1.0 \, 10^{-5}$ at $t=400$.}
\end{figure}
%  Figure 16
\begin{figure}
\epsfxsize=8.5cm
\epsfbox{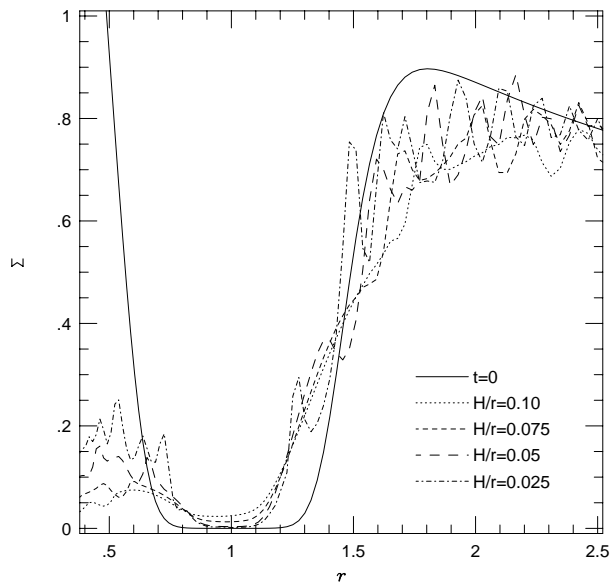}
  \caption{Radial surface density distribution at $\varphi=0$ in
opposition to the planet for various disc temperatures for $t=400$.}
\end{figure}
Aside from the mass accretion rate onto the planet, the physical
extent of the gap is important as it determines observational properties
of protostellar discs.
The radial size of the gap is determined by the mass of the planet
and the properties (viscosity, scale height) of the disc. Keeping the
temperature in the disc fixed ($H/r=0.05$) and varying the
viscosity by a factor of 4 from the standard value we find that the
gap size varies only very slightly with viscosity (Fig.~15).
The data in Fig.~15 are taken at the angle $\varphi = 0$ opposite of the
planet. The inner gap size at the radial location $r=1$ of the planet
is decreasing with increasing viscosity, while the outer part at
about $r=1.55$ is increasing with larger $\nu$. The location of the maxima,
i.e the location of the spirals is not visibly affected by the magnitude
of the viscosity. The range of $\nu$ covers $2.5 \, 10^{-6}$
to $4  \, 10^{-5}$, i.e. a factor of $16$. The first, innermost
bump at $r=1.4$ is created by the secondary spiral (cf. Fig.~2), and then
primary and secondary spirals interchange. A higher viscosity causes a
stronger mass accretion from the outside, and hence a lowering of
the density in the outer parts. The inner region ($r < 1$) is strongly
influenced by mass loss through the inner boundary.
A variation of the non-physical model parameter (artificial viscosity
resolution etc.) as presented in Fig.~6 does not alter
the size and shape of the gap and spirals at all.
%  Figure 17
\begin{figure*}
\epsfxsize=\textwidth
\epsfbox{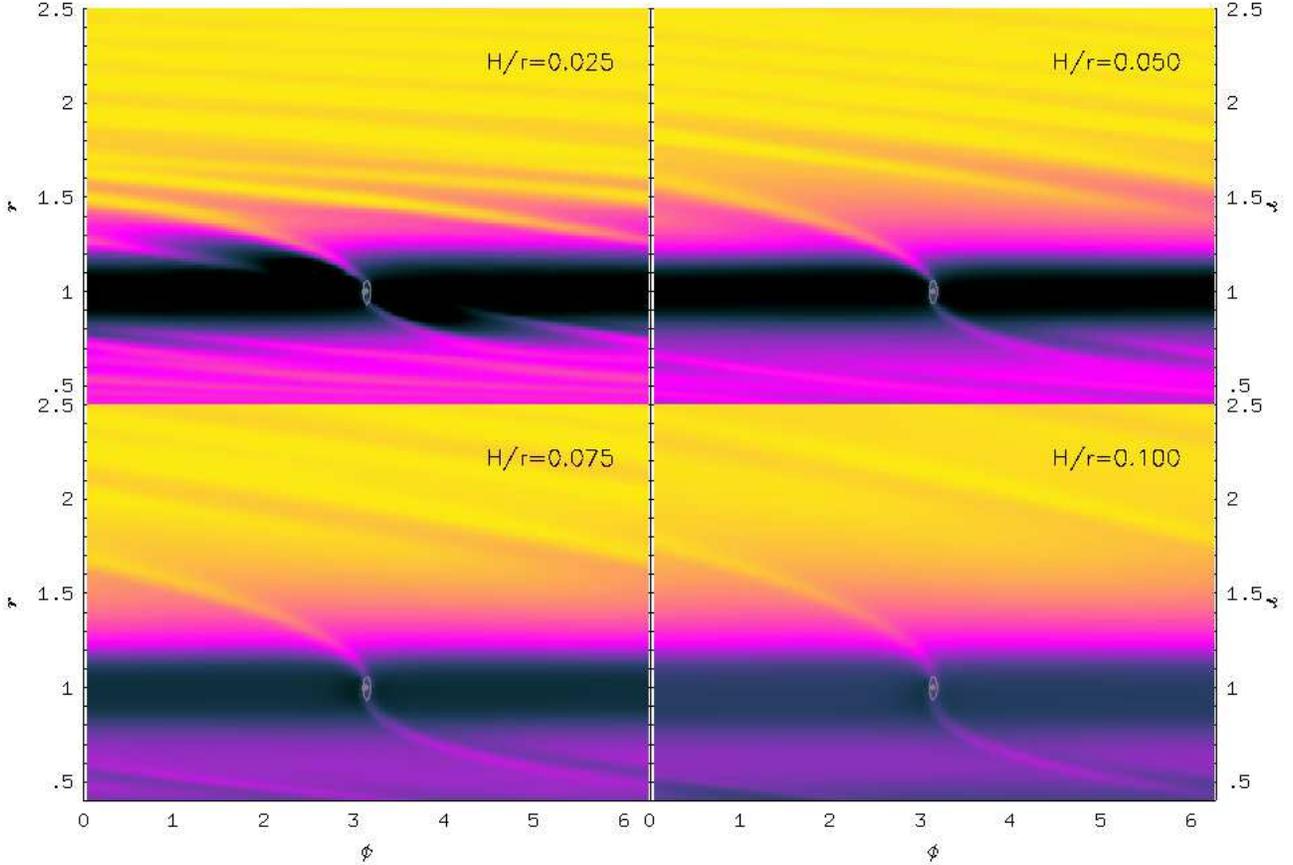}
  \caption{Surface density distribution for four different values
of $H/r$ at $t=400$. The density is scaled as $\Sigma^{1/4}$ between
$4.2 \, 10^{-3}$ and $0.90$.
}
\end{figure*}

On the contrary, a variation of the vertical disc thickness has a
much bigger effect on the structure of gap and spirals (Fig.~16).  
The smaller the temperature the narrower is the gap and the spirals
are more tightly bound. The density enhancement in the spirals
above the surrounding is also more pronounced for lower $H/r$,
as can be seen most clearly from the coolest disc (dashed-dotted line).
At higher temperatures the secondary spiral
disappears. In the plot it is hardly visible anymore for $H/r= 0.075$
(short dashed) and not at all for $H/r=0.1$ (dotted).
The variation of the tightness of the spirals as a function of $H/r$ is
clearly visible in Fig.~17, where gray scale contour plots of the surface
density are displayed for the four different values of $H/R$ (but fixed
$\nu = 10^{-5}$) at the same time of $400$ orbits.
A smaller $H/r$ value clearly leads to a more pronounced gap
with a lower density within the gap region.
%
%  Figure 18
\begin{figure*}
\epsfxsize=\textwidth
\epsfbox{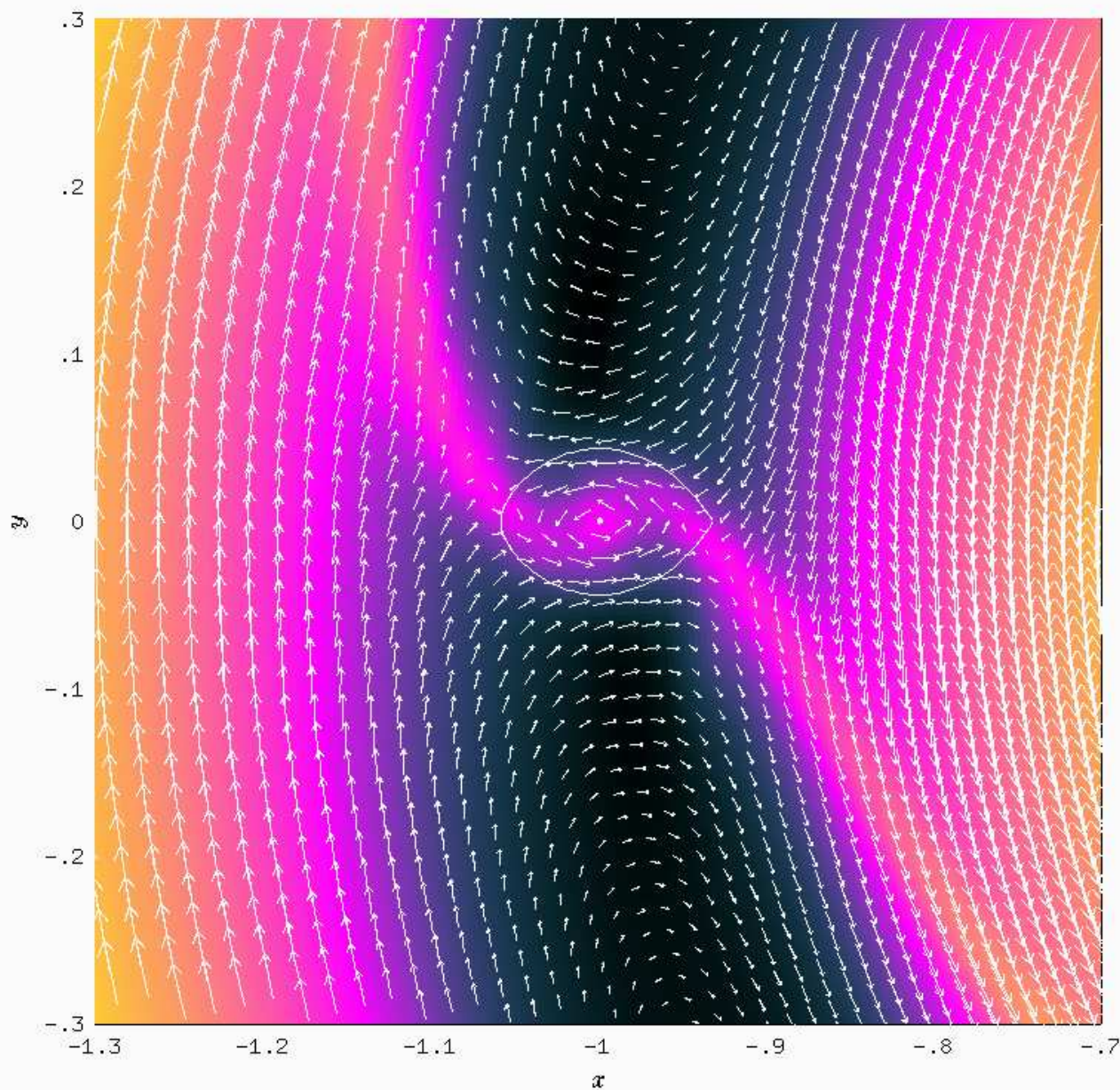}
  \caption{Flow field in the vicinity of the planet in a reference
 frame corotating with the planet. The length of the velocity vectors
  scale logarithmically.}
\end{figure*}
\subsection{The flow in the vicinity of the planet}
The structure of the flow near the protoplanet can only be studied by
using higher resolution models. Here we used a model (2q16) with
$R\subscr{min}=0.4$, $R\subscr{max}=2.5$ and a resolution of $128 \times 442$
gridcells where the radial spacing is logarithmic such that
each gridcell has an approximately square shape in $x-y$ in the
corresponding coordinates. Caused by the higher resolution it was not possible
to follow the whole evolution over several hundred orbits, instead we ran the
model upto $t=100$. As seen above, the flow field at this time has become
approximately stationary and represents quite
accurately the flow at later times.

We would like to point
out that from the flow field it is apparent that the mass accretion
does not occur along the spiral arms (which may be guessed by purely looking
at the density graphs in the previous plots). 
These spiral arms are in fact trailing shock waves. 
The mass supplied to the Roche lobe of the planet comes from material
following the stream lines of the horseshoe orbits in the vicinity of
the planet. From Fig.~18 we notice that some material follows
orbits that enter the Roche lobe
from regions with are lying outside ($r>a$) and inside ($r<a$)
of the planet. This matter is then allowed to be accreted by the planet.
As mentioned above, the accretion rate from inside
is about 3/5 of the one from the outside when the inner boundary is closed
to mass flow. The inner rate is reduced significantly however by the open
inner boundary, which is why we displayed primarily only the value from the
outside. 

The flow field (Fig.~18) near the planet (the central
white dot), whose Roche lobe
is indicated by the solid line, shows that the mass and consequently
angular momentum is accreted in such a way to induce a prograde rotation
of the protoplanet, as it is observed in the solar system for all
massive planets with the exception of Uranus.
The numerical resolution within the Roche lobe is still not very high,
but one can notice nevertheless that the matter orbits the planet.
Whether this circulation around the planet (the {\it proto-Jovian disc})
is Keplerian cannot be determined
from the model, since the smoothing length modifies the potential within
the Roche lobe. Mass does not really accumulate within the Roche lobe
as it is taken out continually to simulate accretion onto the central
planet.

Finally, we mention that a comparison model (with standard resolution)
where no mass accretion onto the planet was taken into account
($f\subscr{red}=0$) produces
a hydrostatic density distribution within the Roche lobe on dynamical
time scales which prevents further mass accumulation. The flow
field, which is similar to Fig.~18, allows in principle for mass
to be transferred across the gap. We find in this case a net mass flow
from radii outside of the planet across the gap to radii
smaller than $a$ of about 1/7 of the standard value
(\ref{macc-2q}) which is substantially smaller than the equilibrium accretion
rate (\ref{mdotacc}).

\section{Conclusions}
We have studied the structure of an accretion disc in a protostellar
environment under the perturbing influence of an embedded protoplanet.
In particular we were interested in the possibility of continued accretion
of mass after the opening of a gap by the planet. Starting from a standard
model with a constant kinematic viscosity coefficient of $\nu = 10^{-5}$
and a vertical thickness of $H/r=0.05$, we first carefully analyzed the
the possible influence of numerical properties such as resolution, artificial
viscosity and rotating coordinates.

Grid resolution and the change
from a corotating coordinate system to an inertial frame do not alter
the physical conclusions, i.e.
have no or very little influence on the mass accretion rate, and the
structure of the gap and trailing shocks.
However, the artificial viscosity has to be treated carefully, and it
turned out that it has to act only on the bulk part of the
viscous stress tensor. This has implications for
calculating similar situations where small deviations from the mean
value of the shear in the underling basic flow are of importance.

The main result of the following calculations has clearly shown that for
all viscosities there is still some accretion onto the planet
taking place, even though the rate is greatly reduced for a very low
viscosity.
The limiting mass of a planet is then determined by the competing
accretion and viscous time scales (see below), and of course by the
available mass reservoir in the surrounding protostellar disc. 
In case of an existing gap the accreted mass originates from material
following streamlines of the horseshoe orbits at the inner and outer edge
of the gap (Fig.~18), where the outside
accretion rate is approximately 1.5 times as large as from the inside
(Eq.~\ref{mdotin}).

There appears to exist a well defined given maximum accretion rate 
(obtained by allowing maximum accretion within the planet's Roche lobe)
depending only on viscosity and temperature in the disc.
For a disc viscosity $\alpha=10^{-3}$ and vertical thickness
$H/r=0.05$ we estimate the timescale for the accumulation
of one Jupiter mass to be of order hundred thousand years.
For a larger(smaller) viscosity and disc thickness this accretion
rate is increasing(decreasing).
The main ingredient required for such a gap accretion is a non vanishing
viscosity within the gap region.
The results with a polytropic equation of state show that in the
case of a viscosity linked to the density (which is very low in the gap)
the accretion rate is smaller. However, one may argue that in a realistic
situation the viscosity, if for example driven by some sort of MHD turbulence,
is also sufficiently large in optically thin regions which would allow
further accretion.

For smaller viscosities $\nu \lsim 10^{-6}$ the mass accretion rate
through the gap onto the planet is markedly reduced (Table~4),
and the corresponding accretion time scale becomes larger than
the viscous evolution time of the disc.
Additional, separate calculations (Bryden et al. 1998) have shown,
that the mass accretion rate onto the planet decreases with larger
$q$, and become for $q = 10^{-2}$ very much longer than $\tau_\nu$.
Hence, the presented calculations show that, for the typical evolution
time scales of protostellar discs ($\tau_\nu \approx 10^{6}$yrs), the final
mass of the planet appears to be in the range $1-10 M\subscr{J}$ consistent
with the observations.

The outlined accretion process onto protoplanets
can only occur in a region of the protoplanetary disc with a sufficiently
large enough mass reservoir, typical for the protostellar disc at a distance
of a few AU from protostar.
As some of the extrasolar planets (51 Peg type planets) have
distances to their stars which are very much smaller than 1 AU, these
planets must then have migrated to their presently observed position
(Lin, Bodenheimer \& Richardson 1996; Ward 1997).

Finally, we would like to point out that our conclusions concerning
the time scale of the accretion process are slightly uncertain,
as they depend on the mass density in the disc surrounding the planet
which scales out of the problem. The numbers stated always refer to
a disc having 0.01 $\Msol$ in the range 1.3 to 20.8 AU. Additionally, the
quoted rates are a lower limit as we only considered here the accumulated mass
from the outer parts of the disc.
Another problem relates to the accretion process within the planet's
Roche lobe. Only if viscous effects in the {\it proto-Jovian} disc
are sufficiently high
to create an $\Mdot$ as large as the supplied rate from the 
{\it protostellar} disc, the maximum
accretion rate (Eq.~\ref{macc-2q}) can be achieved. The evolution of the
proto-Jovian disc is an outstanding problem which needs to be addressed.
The masses of the
extrasolar planets are also lower limits as they still include the uncertain
inclination of the orbit to the line of sight of the observations.
Statistically one may expect them to be a factor 1.3 higher.

Further problems, such as the details of the flow in the vicinity of the
planet, the angular momentum transfer, the influence of a non-vanishing
eccentricity  or orbital inclination of the planet,
the modeling of energy transfer in the disc,
and other issues are beyond the scope of present study. Some of these
future investigations will even require fully three
dimensional calculations.
\section*{Acknowledgments}
I would like to thank Dr. P. Artymowicz for the very detailed and thorough
discussions on this problem. Additionally, I would like thank the
Stockholm Observatory for the kind hospitality during a visit, 
where part of this work was initiated. I am also grateful to the referee
D. Lin for very helpful comments during the refereeing process.
Some of the calculations for this work were
performed on a Cray J90 at the HLRZ J\"ulich.
Computational resources of the Max-Planck Institute for Astronomy
in Heidelberg were also available and are gratefully acknowledged.
This work was supported by the Max-Planck-Gesellschaft,
Grant No. 02160-361-TG74.

\end{document}